\documentclass[times,astrosymb,twocolumn]{aastex631}

\usepackage{natbib}
\usepackage{rotating}
\usepackage{hyperref}
\graphicspath{{./}{Figures/}}
\usepackage{romannum}
\AtBeginDocument{\pagenumbering{arabic}}

\usepackage{graphics,graphicx}
\usepackage{footnote}

\newcommand{\uat}[2]{\href{http://astrothesaurus.org/uat/#2}{#1 (#2)}}
\newcommand{\shortname}{J1429$+$5447}
\newcommand{\longname}{\objectname{CFHQS J142952+544717}}
\newcommand{\cii}{[C\ensuremath{\,\textsc{ii}}]}
\newcommand{\mgii}{Mg\ensuremath{\,\textsc{ii}}}
\newcommand{\RNum}[1]{\uppercase\expandafter{\romannumeral #1\relax}}

\shorttitle{NuSTAR observations of J1429$+$5447}
\shortauthors{Marcotulli et al.}

\graphicspath{{./}{Figures/}}

\begin{document}

\title{NuSTAR observations of a varying-flux quasar in the Epoch of Reionization}

\correspondingauthor{Lea Marcotulli, Thomas Connor}
\email{lea.marcotulli@yale.edu, thomas.connor@cfa.harvard.edu}

\author[0000-0002-8472-3649]{Lea Marcotulli}
\altaffiliation{NHFP Einstein Fellow}
\affil{Yale Center for Astronomy \& Astrophysics, 52 Hillhouse Avenue, New Haven, CT 06511, USA}
\affil{Department of Physics, Yale University, P.O. Box 208120, New Haven, CT 06520, USA}

\author[0000-0002-7898-7664]{Thomas Connor}
\affiliation{Center for Astrophysics $\vert$\ Harvard\ \&\ Smithsonian, 60 Garden St., Cambridge, MA 02138, USA}
\affiliation{Jet Propulsion Laboratory, California Institute of Technology, 4800 Oak Grove Drive, Pasadena, CA 91109, USA}

\author[0000-0002-2931-7824]{Eduardo Ba\~nados}
 \affiliation{Max-Planck-Institut f\"ur Astronomie, K\"onigstuhl 17, D-69117 Heidelberg, Germany}
 \author[0000-0001-9379-4716]{Peter G. Boorman}
 \affiliation{Cahill Center for Astrophysics, California Institute of Technology, 1216 East California Boulevard, Pasadena, CA 91125, USA}
 \author[0000-0002-0786-7307]{Giulia Migliori}
 \affiliation{INAF Istituto di Radioastronomia, via Gobetti 101, 40129 Bologna, Italy}

\author[0000-0002-1984-2932]{Brian W. Grefenstette}
\affiliation{Cahill Center for Astrophysics, California Institute of Technology, 1216 East California Boulevard, Pasadena, CA 91125, USA}
 \author[0000-0003-3168-5922]{Emmanuel Momjian}
 \affiliation{National Radio Astronomy Observatory, P.O. Box O, Socorro, NM 87801, USA}
\author[0000-0002-0905-7375]{Aneta Siemiginowska}
 \affil{Center for Astrophysics $\vert$\ Harvard\ \&\ Smithsonian, 60 Garden St., Cambridge, MA 02138, USA}
\author[0000-0003-2686-9241]{Daniel Stern}
 \affiliation{Jet Propulsion Laboratory, California Institute of Technology, 4800 Oak Grove Drive, Pasadena, CA 91109, USA}%

 \author[0000-0003-4747-4484]{Silvia Belladitta}
 \affiliation{Max-Planck-Institut f\"ur Astronomie, K\"onigstuhl 17, D-69117 Heidelberg, Germany} 
 \affiliation{INAF-Osservatorio di Astrofisica e Scienza dello Spazio di Bologna, Via Piero Gobetti 93/3, 40129 Bologna, Italy}
 \author[0000-0002-4377-0174]{C.~C.~Cheung}
 \affiliation{Space Science Division, Naval Research Laboratory, Washington, DC 20375, USA}
 \author[0000-0002-9378-4072]{Andrew Fabian}
 \affil{Institute of Astronomy, Madingley Road, Cambridge, CB3 0HA, UK}
 \author[0000-0002-7220-397X]{Yana Khusanova}
 \affiliation{Max-Planck-Institut f\"ur Astronomie, K\"onigstuhl 17, D-69117 Heidelberg, Germany}
 \author[0000-0002-5941-5214]{Chiara Mazzucchelli}
 \affiliation{Instituto de Estudios Astrof\'{\i}sicos, Facultad de Ingenier\'{\i}a y Ciencias, Universidad Diego Portales, Avenida Ejercito Libertador 441, Santiago, Chile.}
\author[0000-0003-2349-9310]{Sof\'ia Rojas-Ruiz}
\affiliation{Department of Physics and Astronomy, University of California, Los Angeles, CA, 90095}
\author[0000-0002-0745-9792]{C. Megan Urry}
 \affil{Yale Center for Astronomy \& Astrophysics, 52 Hillhouse Avenue, New Haven, CT 06511, USA}
 \affil{Department of Physics, Yale University, P.O. Box 208120, New Haven, CT 06520, USA}

\begin{abstract}

With enough X-ray flux to be detected in a 160 s scan by \textit{SRG}/eROSITA, the $z=6.19$ quasar \longname\ is, by far, the most luminous X-ray source known at $z>6$. We present deep (245 ks) \textit{NuSTAR} observations of this source; with ${\sim}180$ net counts in the combined observations, \longname\ is the most distant object ever observed by the observatory. Fortuitously, this source was independently observed by \textit{Chandra} ${\sim}110$ days earlier, enabling the identification of two nearby ($30''$ and $45''$ away), fainter X-ray sources. We jointly fit both \textit{Chandra} and \textit{NuSTAR} observations\textemdash self-consistently including interloper sources\textemdash
and find that, to greater than 90\% confidence, the observed $3-7\,\rm keV$ flux varied by a factor of $\sim2.6$ during that period, corresponding to approximately two weeks in the quasar rest-frame. This brightening is one the most extreme instances of statistically significant X-ray variability seen in the Epoch of Reionization. We discuss possible scenarios that could produce such rapid change, including X-ray emission from jets too faint at radio frequencies to be observed.
\end{abstract}


\section{Introduction} \label{sec:intro}
 
\begin{figure*}[t]
\centering
    \includegraphics{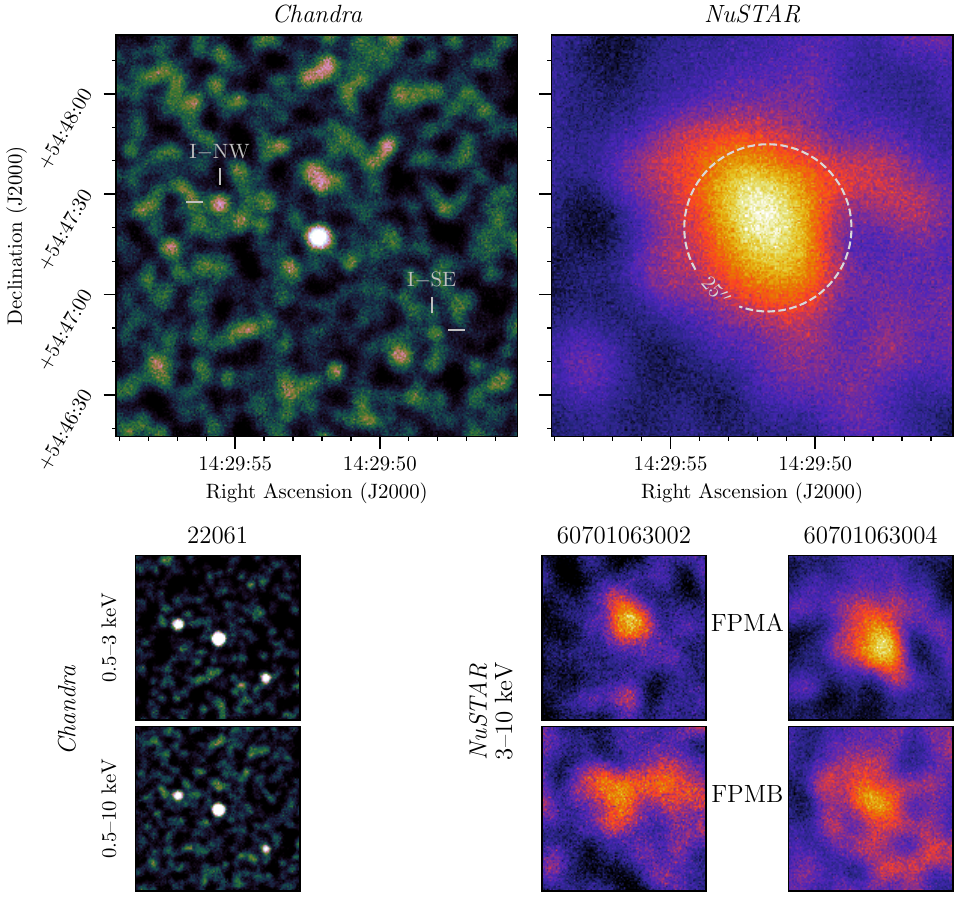}
      \caption{\textit{Chandra} (left) and \textit{NuSTAR} (right) X-ray observations of \shortname. All panels show the same $2^\prime \times 2^\prime$ region, and the sets of images are smoothed by Gaussian kernels of radius $2^{\prime\prime}$ (\textit{Chandra}) and $8^{\prime\prime}$ (\textit{NuSTAR}). \textbf{Top:} $3-10\,\rm keV$ images of the single \textit{Chandra} observation and the combined \textit{NuSTAR} observation. Reticles on the \textit{Chandra} image mark the positions of I-SW and I-NE. A circular aperture of radius $25^{\prime\prime}$ is drawn around the centroid of the \textit{NuSTAR} flux. \textbf{Bottom Left:} \textit{Chandra} observations in a soft 0.5\textendash3.0 keV band (upper) and broad 0.5\textendash10.0 keV band (lower). The two interloper sources are clearly seen in these two bands, despite not contributing significant flux to the $3-10\,\rm keV$ band. \textbf{Bottom Right}: \textit{NuSTAR} 3\textendash10 keV images of \shortname\ from observation 60701063002 (N02, left) and 60701063004 (N04, right) with FPMA (top) and FPMB (bottom). \shortname\ is clearly detected in the combined image as well as the individual exposures, albeit less significantly in the FPMB images owing to the higher background.
      }
      \label{fig:sky_images}
\end{figure*}
 
One of the more exciting revelations of the early Universe has been the discovery of supermassive black holes (SMBHs) with masses in excess of $10^9\ {\rm M}_\odot$ well into the first billion years after the Big Bang \citep[$z \gtrsim 6$;][]{2023ARA&A..61..373F}. Such masses are comparable to the upper reaches of mass found at lower redshift \citep[e.g.,][]{2011ApJS..194...45S}, implying that these black holes have already reached maturity despite their young age. This insight raises a critical question, however, of what mechanism produced these black holes originally \citep{2021NatRP...3..732V}. Seeding hypotheses can generally be classified into either light seeds ($M_{\rm BH}{\lesssim}10^2\ {\rm M}_\odot$, e.g., \citealp{Bromm_1999,2012ApJ...756L..19W}) or heavy seeds ($M_{\rm BH}{\gtrsim}10^4\ {\rm M}_\odot$, e.g., \citealp{Lodato_2006,2019Natur.566...85W}); the latter tend to require more complex and fine-tuned models to produce \citep{2016MNRAS.463..529H, 2020MNRAS.492.4917L}, but the former can only generate the observed SMBHs with faster-than-Eddington growth \citep{2021ApJ...907L...1W}. As such, the current path toward understanding these first SMBHs lies in understanding their conditions and relating these insights to models of their growth. 

\begin{deluxetable*}{llrrrrr}
\tablecaption{X-ray Observations}
\label{tab:summary}
\tablewidth{0pt}
\tablehead{
\colhead{Observatory} & \colhead{Instrument} & \colhead{ObsID} & \colhead{Effective Exposure} & \colhead{Start Date} & \colhead{CR$_{3-5\,\rm keV}$\tablenotemark{a}} & \colhead{CR$_{5-10\,\rm keV}$\tablenotemark{a}} \\
\colhead{} & \colhead{} & \colhead{} & \colhead{(ks)} & \colhead{(YYYY-MM-DD)}& \colhead{$({\rm cts~ks^{-1}})$} & \colhead{$({\rm cts~ks^{-1}})$} }
\startdata
\textit{Chandra} & ACIS-S &  22061 & 30.56 & 2021 Aug 03 & $0.446\pm0.127$ & $0.214\pm0.124$\\
\textit{NuSTAR} & FPMA &  60701063002 (N02) & 78.3 & 2021 Nov 20 & $0.184\pm0.067$ & $0.235\pm0.071$ \\
\textit{NuSTAR} & FPMB &  60701063002 (N02) & 77.6 & 2021 Nov 20 & $0.138\pm0.074$& $0.149\pm0.085$ \\
\textit{NuSTAR} & FPMA &  60701063004 (N04) & 166.4 & 2021 Nov 22 & $0.208\pm0.046$ & $0.263\pm0.055$ \\
\textit{NuSTAR} & FPMB &  60701063004 (N04) & 164.8 & 2021 Nov 22 & $0.117\pm0.047$ & $0.167\pm0.057$  \\
\enddata
\tablenotetext{a}{Count rates are of \shortname\ and are background-subtracted within apertures of radius $5^{\prime\prime}$ (\textit{Chandra}) and $25^{\prime\prime}$ (\textit{NuSTAR}).}
\end{deluxetable*}

Chief among these efforts is discovering more of these objects\textemdash more distant, more massive, and more complete samples. To date, hundreds of SMBHs have been found in the Epoch of Reionization \citep[e.g.][]{2016ApJS..227...11B, 2023ApJS..265...29B, 2019ApJ...884...30W, 2022ApJS..259...18M, 2023ARA&A..61..373F}, with billion Solar mass black holes confirmed out to $z{\gtrsim}7.5$ \citep{2018Natur.553..473B, 2020ApJ...897L..14Y, 2021ApJ...907L...1W} and spectroscopic mass measurements to $z{\sim}10$ \citep[$M_{\rm BH}\approx 10^{6\text{\textendash}7}\ {\rm M}_\odot$, e.g.,][]{2023ApJ...953L..29L, 2024Natur.627...59M}. Detailed studies of these objects have provided critical information about their growth, including that the stellar populations of their host galaxies are undermassive \citep{2023ApJ...957L...3P}, that broad-line region metallicity is established by $z{\sim}7.5$ \citep[e.g.,][]{2020ApJ...898..105O}, and that the first SMBHs reside in massive halos \citep[e.g.,][]{2022ApJ...941..106F}. However, some of the strongest insights into the ongoing accretion come from X-ray observations, which directly trace the accretion ongoing in the Active Galactic Nucleus (AGN) powering the emission.


Owing to both the significant luminosity distances and the effects of redshifting steep power-law spectra \citep{2020ApJ...900..189C}, even luminous quasars from the Epoch of Reonization often yield limited photon statistics \citep[e.g.,][]{2018ApJ...856L..25B}, and insights into these populations sometimes rely on as few as three photons \citep{2019A&A...628L...6V, 2019ApJ...887..171C}. To that end, high-redshift X-ray astronomy is heavily dependent on extrapolating from lower-redshift insights, such as the observed correlation between power-law spectral slope $\Gamma$ and accretion rate \citep{2013MNRAS.433.2485B, 2017MNRAS.470..800T, 2018MNRAS.480.1819R} or the scaling between UV luminosity and UV\textendash to\textendash X-ray spectral slope \citep{2016ApJ...819..154L}. 
At this epoch, however, to accurately characterize higher-energy relations such as that between coronal temperature and compactness, broader bandwidth is needed than what is currently achievable with instruments such as \textit{Chandra} and \textit{XMM-Newton}\textemdash despite their power in characterizing accretion regions \citep{2015MNRAS.451.4375F, 2019ApJ...875L..20L}.


Much of the work on the X-ray properties of the first quasars therefore comes from stacking analyses and population statistics \citep{2017A&A...603A.128N, Vito_2019, 2021MNRAS.504.2767L, 2021ApJ...908...53W, Zappacosta_2023}. Buttressing these works is the benefit of redshift on observed times; for a source at redshift $z$, the time elapsed in the observer frame, $t_{\rm obs}$, is related to the time in the rest-frame ($t_{\rm rest}$) as $t_{\rm obs} = t_{\rm rest} (1 + z)$. Thus, observations spread across years only sample source variations on time scales of months, and even the longest continuous X-ray observations cover just a few hours of rest-frame variability.  Nevertheless, we have already witnessed a few cases of fast \citep[e.g.~$t_{\rm rest}\sim250s$;][]{Moretti_2021} and extreme \citep[flux decrease of a factor of $>7$ in $t_{\rm rest}\sim 115$ days;][]{2022A&A...663A.159V} X-ray variability of $z>6$ quasars, shedding a light onto the physical processes powering these AGN. 

It is this context that greeted the X-ray detection of \longname\ (hereafter \shortname), a $z=6.19$ \citep{2022AA...664A..39K} quasar first discovered by \citet{2010AJ....139..906W}. Powered by a black hole with a \mgii-derived mass measurement of $\log({\rm M}_{\rm BH} / {\rm M}_\odot) = 9.26 \pm 	0.37$ \citep{2019ApJ...873...35S}, the quasar is radio-loud but shows no extended radio structure beyond 100 pc in projection \citep{2011A&A...531L...5F}. The source was detected by \textit{SRG}/eROSITA\footnote{\shortname\ is located in the Eastern Galactic hemisphere, the side where analysis of eROSITA data is led by the Russian consortium.} in 160 s \citep{2020MNRAS.497.1842M}; follow-up \textit{XMM-Newton} observations confirmed this detection, and established that \shortname, with a $1.4-72\,\rm keV$ (rest-frame) luminosity of ${\rm L}_{\rm X} = 5.5^{+0.8}_{-0.6}~\times~10^{46}\ {\rm erg}\ {\rm s}^{-1}$, is far and away the most X-ray luminous object in the Epoch of Reionization \citep{2021MNRAS.504..576M, 2023MNRAS.524.1087M}, surpassing runners-up SDSS J010013.02+280225.8 \citep[$z=6.3$,][]{2021ApJ...922L..24C} and PSO J030947.49+271757.31 \citep[$z=6.1$,][]{Belladitta_2020} by around an order of magnitude.

Not only is \shortname\ X-ray luminous, it is also radio-loud \citep[radio-loudness parameter $R= f'_{\nu,\rm 5 GHz}/f'_{\nu,\rm 4400\ \text{\normalfont \AA}}=109\pm9$ and
$L_{\rm 5 GHz} {=} 10^{9.6}\ L_\odot$;][]{Kellermann_1989, 2015ApJ...804..118B}, a feature typically associated with strong jets. Despite \cii\ observations \citep{2022AA...664A..39K} showing evidence for an AGN-driven outflow, \citet{2011A&A...531L...5F} found that the size of the resolved radio emission is only around 100 pc in projection and measured a 2-point (1.6\textendash 5 GHz) radio 
spectral index of $\alpha \sim {-1.0}$ (see also \citealt{2017MNRAS.467.2039C}); together, along with the relatively low rest-frame brightness temperature, these observations disfavor a blazar interpretation (i.e. the AGN hosts a relativistic jet pointed close to our line of sight, \citealp{Urry_1995}).
Conversely, recent CO spectral line energy distribution (SLED) observations seem to imply that the host galaxy could be strongly impacted by its jets (Khusanova et al., submitted, although cf. \citealp{2024ApJ...962..119L}). The lack of observed large ($>$kpc scale, \citealp{Mingo_2019}) radio jets may be because jet emission is dominated by cosmic microwave background (CMB) photons being inverse Compton scattered (IC/CMB; e.g., \citealp{2014MNRAS.438.2694G, Ghisellini_2015}). Such X-ray\textendash dominated jets could potentially be tracers of needed rapid accretion \citep{2008MNRAS.386..989J}, but detections of extended X-ray jets at these redshifts have been so-far limited \citep{2021ApJ...911..120C,Ighina_2022} and may remain so until the next generation of X-ray satellites is launched \citep{2024Univ...10..227C, Marcotulli_2024}. 


\shortname, at the extremes of the early Universe, is thus mystifying: radio-loud and X-ray luminous, but with neither extended jets nor the hallmarks of being a blazar. Yet its brightness makes it accessible at very hard energies ($E'_{\rm rest-frame}>50\,\rm keV$), where AGN emission can be more easily isolated. We therefore obtained \textit{NuSTAR} observations of the quasar to constrain possible X-ray emission scenarios that could be producing such extreme luminosity. In this paper we describe these observations, as well as the role of short-term variability from \shortname\ in comparison to independent \textit{Chandra} observations and the potential origins\textemdash and impacts\textemdash of that variability. 
 
Throughout this work, we use a flat cosmology with $H_0 = 70\,\textrm{km\,s}^{-1}\,\textrm{Mpc}^{-1}$, $\Omega_M = 0.3$, and $\Omega_\Lambda = 0.7$. We adopt a redshift for \shortname\ of $z=6.19$, as reported from \cii\ line measurements by \citet{2022AA...664A..39K}; at this redshift, the scale is $ 5.61\ {\rm kpc}\, \textrm{arcsec}^{-1}$. Per \citet{2016A&A...594A.116H}, we assume a Galactic neutral hydrogen column density of $N_{\rm H}= 1.15\times 10^{20}\,\textrm{cm}^{-2}$ in the direction of \shortname. All distances given are in proper distances and errors are reported at the 1$\sigma$ (68\%) confidence level unless otherwise stated. For reported radio spectral slopes, we follow the convention of $S_\nu \propto \nu^\alpha$; we instead use $\Gamma$ for the reported X-ray photon indices following the convention $dN/dE\propto E^{-\Gamma}$ (where $\Gamma=\alpha-1$). 

\section{Observations and Data Reduction}\label{sec:Observations}

In this work we consider two sets of X-ray observations: archival \textit{Chandra} imaging and new \textit{NuSTAR} data. The summary of these observations, including the background subtracted count rates, can be found in Table~\ref{tab:summary}. There are two X-ray interlopers in this field which impact these observations, which are discussed below. The coordinates and redshifts of these interlopers and the quasar \shortname\ are given in Table~\ref{tab:xraysources}.

\subsection{Interloper Sources}
There are two known X-ray interlopers in this field. The first, identified with \textit{XMM-Newton} by \citet{2021MNRAS.504..576M}, is approximately $45^{\prime\prime}$ southwest of the quasar, while the second, identified in higher-resolution \textit{Chandra} imaging by \citet{2023MNRAS.524.1087M}, is roughly $30^{\prime\prime}$ to the quasar's northeast. For clarity, we refer to these interloping sources as ``I-SW'' and ``I-NE,'' respectively, throughout this work. Using the robust \textit{Chandra} astrometry, we cross-matched the positions of these sources with the \textit{Hubble} Source Catalog \citep{2016AJ....151..134W}, the  Canada-France-Hawaii Telescope Legacy Survey \citep[CFHTLS,][]{2012SPIE.8448E..0MC}, and the Legacy Imaging Surveys photometric redshift catalog of \citet{2022MNRAS.512.3662D}. Neither source has a radio detection within $10^{\prime\prime}$ indexed by VizieR \citep{2000A&AS..143...23O}.

\begin{deluxetable}{lllr}
\tablecaption{X-ray Sources}
\label{tab:xraysources}
\tablewidth{0pt}
\tablehead{
\colhead{Name} & \colhead{R.A.} & \colhead{Dec.} & \colhead{Redshift} }
\startdata
\shortname & 14:29:52.18 & +54:47:17.6 & $6.190\pm0.004$\\
I-SW & 14:29:48.21 & +54:46:49.1  & $1.23\pm0.50$\\
I-NE & 14:29:55.54 & +54:47:27.8 & $1.02\pm0.13$\\
\enddata
\tablerefs{\shortname's redshift is from a fit to the [\ion{C}{2}]$\lambda 158~\mu{\rm m}$ line \citep{2022AA...664A..39K}, while the interlopers' redshifts are photometric \citep{2022MNRAS.512.3662D}.}
\end{deluxetable}

I-SW has NIR magnitude of  ${\rm F}105{\rm W}=22.4$, CFHTLS color of $i^{\prime} - z^{\prime}\approx 0.2$, and photometric redshift $1.23\pm0.50$. I-NE has magnitude ${\rm F}105{\rm W}=21.5$, color $i^{\prime} - z^{\prime}\approx 0.9$, and photometric redshift $1.02\pm0.13$. In contrast, the CFHTLS color of \shortname\ is $i^{\prime} - z^{\prime}\approx 2.5$, owing to the presence of the Lyman $\alpha$ break around 8740 \AA, significantly suppressing flux in the $i^\prime$ band. Because the tell-tale sign of the Lyman $\alpha$ break is not present, and in conjunction with the photometric redshifts, we conclude that these are foreground interloper sources, unrelated to the quasar or its system.

\subsection{Chandra}

\shortname\ was observed by the \textit{Chandra} X-ray Observatory for 30.56 ks on 2021 Aug 03 using the Advanced CCD Imaging Spectrometer \citep[ACIS;][]{2003SPIE.4851...28G}. In the rest-frame of the quasar, these observations took place approximately 15 days before the \textit{NuSTAR} observations. The quasar was positioned on the back-illuminated S3 chip, and ACIS was operated in the Timed Exposure (TE) mode with Very Faint (VF) telemetry saturation limits. These data, consisting of a single observation, are contained in \dataset[CDC 308]{https://doi.org/10.25574/cdc.308} and were previously published by \citet{2023MNRAS.524.1087M}.

We reduce the same observation using the \textit{Chandra} Interactive Analysis of Observations software package \citep[CIAO;][]{2006SPIE.6270E..1VF} v4.14 with CALDB v4.15. We define circular source regions of $5''$ radius centered, respectively, at the position of the quasar, I-NE, and I-SW. Annulus background regions of $7''$ inner and $23''$ outer radii centered at the source locations are selected. The spectra for all three sources are obtained with the \texttt{specextract} function, and the spectral files are binned with the optimal binning scheme of \texttt{ftgrouppha} \citep{Kaastra_2016}. We fit our spectra with BXA (see Section~\ref{sec:chandra}) and compare our results to those of \citet{2023MNRAS.524.1087M}.
Modulo uncertainties, neither I-NE or I-SW is brighter than \shortname, nor do they have shallower spectra; as such, and as evidenced by Figure \ref{fig:sky_images}, their contribution to the hard band probed by \textit{NuSTAR} should be negligible---although we nevertheless include both sources in our fits.

\subsection{NuSTAR}

We observed \shortname\ with \textit{NuSTAR} \citep{2013ApJ...770..103H} as part of program 7291. These observations were split into two segments: an effective exposure time of 78.3 ks (FPMA) and 77.6 ks (FPMB) starting on 2021 Nov 20 (ObsID: 60701063002, hereafter referred to as N02) and 166.4 ks (FPMA) and 164.8 ks (FPMB) starting on 2021 Nov 22 (ObsID: 60701063004, hereafter referred to as N04). Although these observations were both preceded and separated by Solar observations in support of a \textit{Parker} Solar Probe perihelion passage, instrumental temperatures were all within typical operating parameters for the duration of the observations and no problems were identified in quality assurance monitoring. 

The \textit{NuSTAR} observations were analysed using the most recent \textit{NuSTAR} calibration database (CALDB) v.~2024-03-25 and HEASOFT v.~6.32.1. The data were pre-processed using the standard \texttt{nupipeline} routine. Although the source is very faint in the full 3\textendash 80 keV \textit{NuSTAR} band, the 3\textendash 10 keV images (Figure~\ref{fig:sky_images}) show a clear excess at the location of the source, in both observations and in both modules. 

Since the \textit{NuSTAR} PSF is large with respect to that of \textit{Chandra} (FWHM of $18''$ vs.~$0.5''$), the emission visible in the \textit{NuSTAR} image may be due to a combination of \shortname, I-SW, and I-NE. Ideally, for our analysis, we would prefer to select a source region that encompasses both the location of the quasar and the interlopers; their measured separations (as can be seen in Figure~\ref{fig:sky_images}) require that a circular region have a radius of at least $49''$ to include all three sources. A radial profile of the \textit{NuSTAR} images generated using the \textit{NuSTAR} General Utilities\footnote{\url{https://github.com/NuSTAR/nustar-gen-utils}} at the emission centroid location does not help us differentiate between one or multiple sources contributing to the \textit{NuSTAR} counts within a $49''$ radius due to the low number of source counts. However, most of the source counts in the \textit{NuSTAR} image are concentrated in a circular region of radius $\sim 25''$, which, if selected as our source region extent, maximizes the signal-to-noise of the data at hand and therefore improves our fit statistics\footnote{The radial profile of the source in the 3\textendash 10 keV energy range shows that the net count rate drops by a factor $\sim1/e$ after $\sim 25''$ from the centroid location, and the signal-to-noise profile does not improve enlarging the source region to $49''$.}.

We therefore extracted spectra using both a signal-optimized circular region ($r=25^{\prime\prime}$) and a more inclusive circular region ($r=49^{\prime\prime}$) that covers the positions of both interloper sources. Both regions were centered on the centroids of the counts in the \textit{NuSTAR} image (which corresponds to the coordinates of \shortname\ to within $7^{\prime\prime}$). Backgrounds were extracted from a circular region of radius $50^{\prime\prime}$ for the smaller source region and $80^{\prime\prime}$ for the larger source region, located away from source contamination but within the same detector area. The spectra of both source and background regions were computed using the \texttt{nuproducts} routine, and consistently the spectral files were binned with the optimal binning scheme of \texttt{ftgrouppha}.

\section{X-ray fitting with BXA}\label{ref:fit_bxa}
\shortname~is clearly detected in both the \textit{Chandra} and \textit{NuSTAR} observations. However, its observed faintness requires our analysis to be fine-tuned to the low-count statistics regime. To properly account for the Poisson nature of the data without background modeling, we minimized the modified C-statistic \citep[W-stat;][]{1979ApJ...228..939C,1979ApJ...230..274W}. We use the Bayesian X-ray Analysis (BXA, v.4.1.1, \citealp{Buchner_2014,Buchner_2016}) software package that links the nested sampling algorithm \texttt{UltraNest} \citep[v. 3.5.7,][]{Buchner_2021} with the X-ray Spectral Analysis software \texttt{XSPEC} \citep{1996ASPC..101...17A,Gordon_2021}. Here we provide a brief overview on how nested sampling works; further, specific details can be found in \citet{Buchner_2023}. 

\begin{deluxetable*}{cccccc}
\tablecaption{Spectral Fits}
\label{tab:spectralfits}
\tablewidth{0pt}
\tablehead{
\multicolumn{6}{c}{\textit{Chandra}} \\
\hline
\colhead{Source} & \colhead{Model} & \colhead{$N_{H,z}$} & \colhead{$\Gamma$} & \colhead{$\log \left( {\rm F}_{0.5 - 10\ {\rm keV}}\right)$} & \colhead{$\log \left( {\rm F}_{3 - 7\ {\rm keV}}\right)$} \\
\colhead{} & \colhead{} & \colhead{($\times 10^{22}\,{\rm cm}^{-2}$)} & \colhead{} & \colhead{[${\rm erg}\ {\rm cm}^{-2}\ {\rm s}^{-1}$]} & \colhead{[${\rm erg}\ {\rm cm}^{-2}\ {\rm s}^{-1}$]}}
\startdata
                     \shortname    &   M1 &  \nodata &  $2.32_{-0.33}^{+0.15}$ & $-13.25_{-0.06}^{+0.04}$ & $-13.92_{-0.11}^{+0.10}$ \\ 
                     I-NE     &   M2 & $19.02_{-8.09}^{+6.89}$ & $3.58_{-0.49}^{+1.54}$ & $-13.93_{-0.12}^{+0.09}$ & $-14.45_{-0.20}^{+0.21}$ \\ 
                     I-SW     &   M1 & \nodata & $2.20_{-0.83}^{+0.41}$ & $-13.92_{-0.12}^{+0.12}$ &  $-14.48_{-0.14}^{+0.29}$ \\
\hline
\vspace{5pt}\\
\multicolumn{6}{c}{\textit{NuSTAR}} \\
\hline
\colhead{Observation} & \colhead{Module} & \colhead{Model} & \colhead{const} & \colhead{$\Gamma$} & \colhead{$\log \left( {\rm F}_{3 - 78\ {\rm keV}}\right)$}\\
\colhead{} & \colhead{} & \colhead{} & \colhead{} & \colhead{} & \colhead{[${\rm erg}\ {\rm cm}^{-2}\ {\rm s}^{-1}$]}\\
\hline
60701063002& FPMA  & M1 &  $1$ (fixed) & $2.34_{-0.73}^{+0.68}$ &  $-13.14_{-0.19}^{+0.18}$ \\
        60701063002& FPMB  & \nodata &  $0.96_{-0.07}^{+0.11}$ & \nodata &  \nodata \\
        60701063004& FPMA  & M1  &  1 (fixed)  &$2.49_{-0.24}^{+0.42}$ &  $-13.08_{-0.10}^{+0.09}$\\
        60701063004& FPMB  & \nodata & $0.88_{-0.08}^{+0.10}$ & \nodata &  \nodata\\
\hline
\vspace{5pt}\\
\multicolumn{6}{c}{Joint \textit{Chandra} and \textit{NuSTAR}} \\
\hline
\colhead{Instrument} & \colhead{Source} & \colhead{Model} & \colhead{$N_{H,z}$} & \colhead{$\Gamma$} & \colhead{$\log \left( {\rm F}_{3 - 7\ {\rm keV}}\right)$}\\
\colhead{} & \colhead{} & \colhead{} & \colhead{($\times 10^{22}\,{\rm cm}^{-2}$)} & \colhead{} & \colhead{[${\rm erg}\ {\rm cm}^{-2}\ {\rm s}^{-1}$]}\\
\hline
\textit{Chandra}      &   \shortname & M1 & \nodata  &  $2.22_{-0.12}^{+0.31}$ & $-13.95_{-0.08}^{+0.11}$  \\ 
                          &   I-NE  & M2 & $16.59_{-5.63}^{+4.78}$ & $3.39_{-0.44}^{+0.48}$ & $-14.35_{-0.16}^{+0.12}$ \\ 
                          &   I-SW  & M1 & \nodata &  $1.73_{-0.48}^{+0.42}$ & $-14.45_{-0.17}^{+0.25}$ \\ 
Co-added FPMA &  \shortname & M1 & \nodata& $2.22_{-0.54}^{+0.60}$ &  $-13.49_{-0.10}^{+0.07}$\\
\enddata
\tablecomments{The values reported in the Table are the mode and the 68\% confidence level obtained from the posterior distribution of the parameters. The reported fluxes are the observed absorbed ones, i.e.~we do not correct for the Galactic (or source) absorption. {\bf \textit{Chandra}:} We separately fit the three sources in the image: the quasar and I-SW with a redshifted powerlaw (M1, Section~\ref{sec:mod}) and I-NE with an absorbed redshifted power-law (M2). {\bf \textit{NuSTAR}:} the \textit{NuSTAR} observations are fit separately with a redshifted power law to test for day time-scale variability (see Section~\ref{sec:day-var}) over the full \textit{NuSTAR} energy band (3\textendash 78 keV) using the spectral products obtained with a $25''$ radius source region. Scaling between FPMA and FPMB was included through multiplicative constants. {\bf Joint:} The \textit{NuSTAR} data are fit as the sum of all three sources, with the interloper spectra fixed to the \textit{Chandra} fits (see Section~\ref{sec:month-time} for details).}
    \label{tab:chan}
\end{deluxetable*}

To each of the fitted parameters of the tested model, we assign a prior distribution of possible values stored as a vector by the nested sampling algorithm. Then the likelihood space of the parameters' combination is explored iteratively, with the lowest likelihood parameter vector being replaced by a new one which has a higher likelihood value. 
This scan terminates when the contribution to the Bayesian evidence integral stabilizes and the likelihood weighted with the volume removed is negligible \citep[see Section 3.4.2 of][]{2023arXiv230905705B}. 
The choice of priors depends on the fitted models and free parameters. In our specific case, where we use power-law-like models to fit our sources, we assign uniform priors to the spectral indices and log-uniform priors to the intrinsic normalizations. When fitting the simultaneous \textit{NuSTAR} observations (see Section~\ref{sec:day-var}), we include a cross-calibration between the two detector modules; this parameter is fixed to 1 for FPMA and assigned a custom log-Gaussian prior with mean zero (i.e. a linear cross-calibration of unity, \citealp{Madsen_2017}) and 0.1 standard deviation for FPMB. When fitting non-simultaneous observations from different instruments, the cross-calibration constants are a binary flag 
used to turn different components of the fit on or off (see Section~\ref{sec:month-time}, and Appendix~\ref{sec:cross_cal_appb} for more details). 

\subsection{Tested models}\label{sec:mod}
For the sake of clarity and reproducibility, we list here the XSPEC spectral model used in the next sections.  
\begin{itemize}
    \item  \textbf{M1}: A redshifted power law\\ \texttt{const*tbabs*zpowerlaw}
    \item  \textbf{M2}: An absorbed redshifted power law\\ \texttt{const*tbabs*ztbabs*zpowerlaw}
    \item  \textbf{M3}: A redshifted broken power law\\  \texttt{const*tbabs*zbknpower}
    \item  \textbf{M4}: A redshifted power law with exponential cutoff\\
    \texttt{const*tbabs*zcutoffpl}
\end{itemize}

The Galactic absorption is frozen at the Galactic value ($N_{\rm H}=1.15 \times 10^{20}\,\rm cm^{-2}$, \citealp{2016A&A...594A.116H}) with the \texttt{tbabs} XSPEC model. The chemical abundance is set to that of \citet[][\texttt{wilm}]{Wilms_2000}.
Throughout the paper, we report the mode of each parameter distribution alongside the highest density interval (HDI) at the 68\% level as our uncertainty, extracted on the posterior distribution for each of the fitted parameters. 
Finally, for every combination of parameters in the posterior distribution, we can derive the posterior distribution of flux and luminosity for every tested model in any desired energy band; the mode and HDI (68\%) of these distributions are quoted in the paper as our source's flux and luminosity.

\section{Results}
\subsection{Chandra}\label{sec:chandra}
Our first objective was to establish X-ray properties for the three sources seen in the \textit{Chandra} observations 
to ensure the differences in reduction and analysis from the work of \citet{2023MNRAS.524.1087M} had no systematic effects. To that end, we fit the existing \textit{Chandra} data in the 0.5\textendash 7.0 keV band for all three sources; for \shortname\ and I-SW we used the redshifted power law (M1, Section~\ref{sec:mod}), while, since I-NE shows hints of an extra absorber \citep{2023MNRAS.524.1087M}, we fit it with a redshifted power law with an intrinsic, redshifted absorber (M2). Adopted redshifts for these sources are given in Table~\ref{tab:xraysources}. The results of these fits are given in Table~\ref{tab:spectralfits}. 


In the case of \shortname~we recover a photon index of $\Gamma=2.32^{+0.15}_{-0.33}$ and an observed unabsorbed flux in the 0.5-10 keV range of $F_{\rm 0.5\text{\textendash} 10\, \rm keV}= 5.6^{+0.6}_{-0.7}\times10^{-14}\,\rm erg~cm^{-2}~s^{-1}$, completely consistent with the results in \citet{2023MNRAS.524.1087M}. The results on I-SW 
also agree within errors with what is reported in \citet[][cf.~Table 1 of their work]{2023MNRAS.524.1087M}. 
We find that for I-NE the column density of the extra absorber is $\sim1.9\times10^{23}\,\rm cm^{-2}$, indicating that the source is intrinsically obscured at soft X-ray energies. The derived photon index, $\Gamma_{\rm I-NE}= 3.39_{-0.44}^{+0.48}$, is quite soft though consistent within $3\sigma$ to the values reported in \citet{2023MNRAS.524.1087M}. The differences in spectral parameters\textemdash mostly limited to absorption\textemdash found over the results of \citet{2023MNRAS.524.1087M} can be explained by the fact that we include the redshift of the interloper, while the earlier analysis did not; the recovered X-ray flux of the interloper is consistent within statistical uncertainties.

\subsection{Independent NuSTAR Observations}\label{sec:day-var}

Although the start times of the two \textit{NuSTAR} observations are separated by two days, there is only a gap of roughly seven hours between the end of N02 and the start of N04. Nevertheless, we fit both observations independently, both to compare the derived properties of our two source extraction regions and to search for rapid variability. For both observations, we fit the observed spectra with a redshifted power law (M1) over the full 3\textendash 78 keV range, with FPMA and FPMB linked within each observation by a freely-varying cross-normalization coefficient.

We separately fit both the $25^{\prime\prime}$ and $49^{\prime\prime}$ radius extraction region spectra, with the results of the former shown in Table~\ref{tab:spectralfits}. When using the larger region, we found no meaningful change in flux or spectral index, but the lower signal-to-noise results in larger statistical uncertainties. As such, we do not use the spectra extracted with the larger regions in further analysis.
Additionally, as shown in Table~\ref{tab:spectralfits}, there is no detectable variability between the two \textit{NuSTAR} observations, either in flux or in spectral shape. Furthermore, the cross-calibration constants between the \textit{NuSTAR} modules are consistent with unity, within errors. For subsequent fits, we opted to co-add the FPMAs from both observations 
with the python wrapper of the FTOOL \texttt{addspec}\footnote{\url{https://github.com/JohannesBuchner/addspec.py}}, binning them with \texttt{ftgrppha} using the optimal binning. 

Moreover, we decided to not include the FPMB observations as the source was located in a region of the \textit{NuSTAR} field of view in which the aperture background contribution is higher in FPMB than the corresponding location in FPMA (see Figure~\ref{fig:sky_images}). The different background levels for the
two FPMs can cause issues when the difference in backgrounds is comparable to the flux of the source  \citep{2014ApJ...792...48W, Nuserendip_2024}. This choice reduces the dimensionality of the parameter space explored by BXA, thus dramatically decreasing computational time, without impacting our findings in a statistically meaningful way. We note that limiting our analysis to {\it NuSTAR} data below 15 keV gives statistically indistinguishable results, as the W-statistic allows us to fit background-dominated spectral regions \citep{2023arXiv230905705B}.

\begin{figure*}[t]
    \centering
    \includegraphics[width=0.8\textwidth]{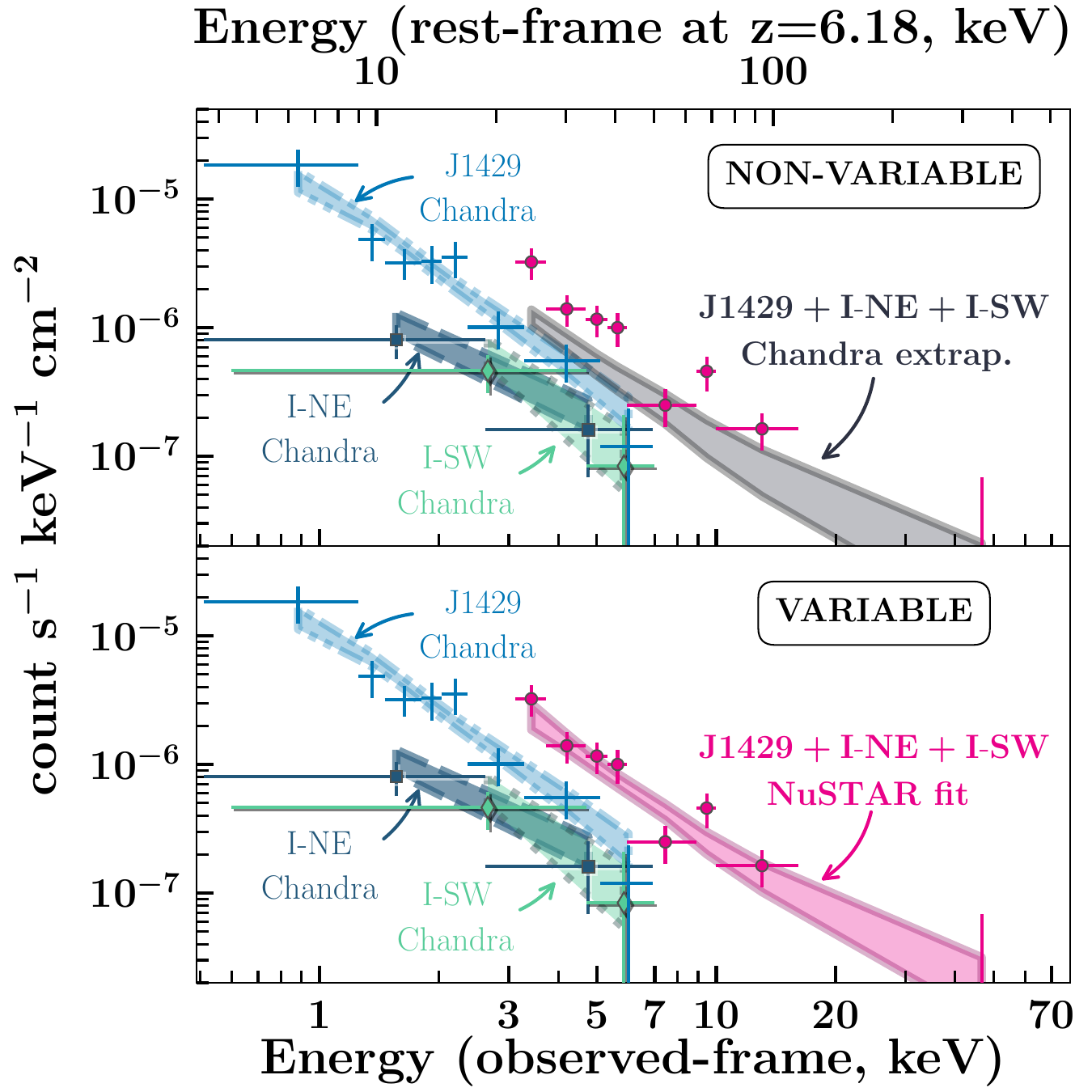}

    \caption{Background-subtracted counts spectra for all three \textit{Chandra} sources (\shortname, light-blue crosses; I-NE, navy squares; I-SW, green diamonds) and the \textit{NuSTAR} data (fuchsia points), after binning each spectrum to have a minimum significance of three per bin for visual clarity. Source counts are not unfolded through the model but are instead normalized by the average effective area in each bin. In both panels, we show the model predictions for the fits to the Chandra data of \shortname\ (light blue, dash-dotted lines), I-NE (navy, dashed line), and I-SW (green shade, dotted line). The shaded bands in the plot represent the 68th percentile range regions of model predictions, as calculated from our posteriors. 
    The \textbf{top} panel shows the \textit{Chandra}-only fits extrapolated (and summed) into the \textit{NuSTAR} passband (gray shade, solid line), which is representative of the scenario in which none of the sources are variable; the \textbf{bottom} panel instead shows the result of our fit in which \shortname\ was allowed to vary in its normalization and photon index between the \textit{NuSTAR} and \textit{Chandra} epochs  (fuchsia shade, solid line),     under the assumption that I-NE and I-SW did not vary in the 4 months (observed frame; $<2$ months rest-frame assuming their photometric redshifts in Table~\ref{tab:xraysources}) between the two observations. The results indicate that the \shortname\ X-ray flux varied by a factor of 2.6 at $>90\%$ CL (see also Figure~\ref{fig:dist_flux}).}
    \label{fig:x-ray-fit}
\end{figure*}
 
\subsection{Combined Chandra and NuSTAR}\label{sec:month-time}

Separated by ${\sim}110$ days in the observed frame, the \textit{Chandra} and \textit{NuSTAR} observations sample any potential variability on rest-frame scales of ${\sim}15$ days \citep{Lewis_2023}. To test for such variability, we performed a simultaneous fit on both data sets, including all three sources in the \textit{Chandra} observations and the single extraction in the co-added \textit{NuSTAR} FPMA observation. We modeled all three sources as for the \textit{Chandra}-only observations: redshifted power laws (M1) for \shortname\ and I-SW and an absorbed redshifted power law (M2) for I-NE. Fits were over the 0.5\textendash7.0 keV (\textit{Chandra}) and 3\textendash 78 keV (\textit{NuSTAR}) ranges. 

In total, the fit was composed of six models: \shortname\, I-NE, and I-SW, fit independently for their \textit{Chandra} spectra and in combination for the \textit{NuSTAR} spectrum. The values for the two interloper sources were linked between observatories, under the assumption that these sources did not meaningfully vary over the observed time difference (see discussion in Section~\ref{sec:disc-int}). Although the interlopers were not centered in the \textit{NuSTAR} aperture, we conservatively included their entire flux in the \textit{NuSTAR} spectra. For \shortname, however, both the normalization and photon index were free to vary independently in both observations. BXA fits return the posterior distribution of parameters for all components, and so, for each set of posteriors, we can extract flux measurements in any desired energy band, although we focus on the overlapping 3\textendash 7 keV range.

\begin{figure*}[t]
    \centering
    \includegraphics[width=0.95\textwidth]{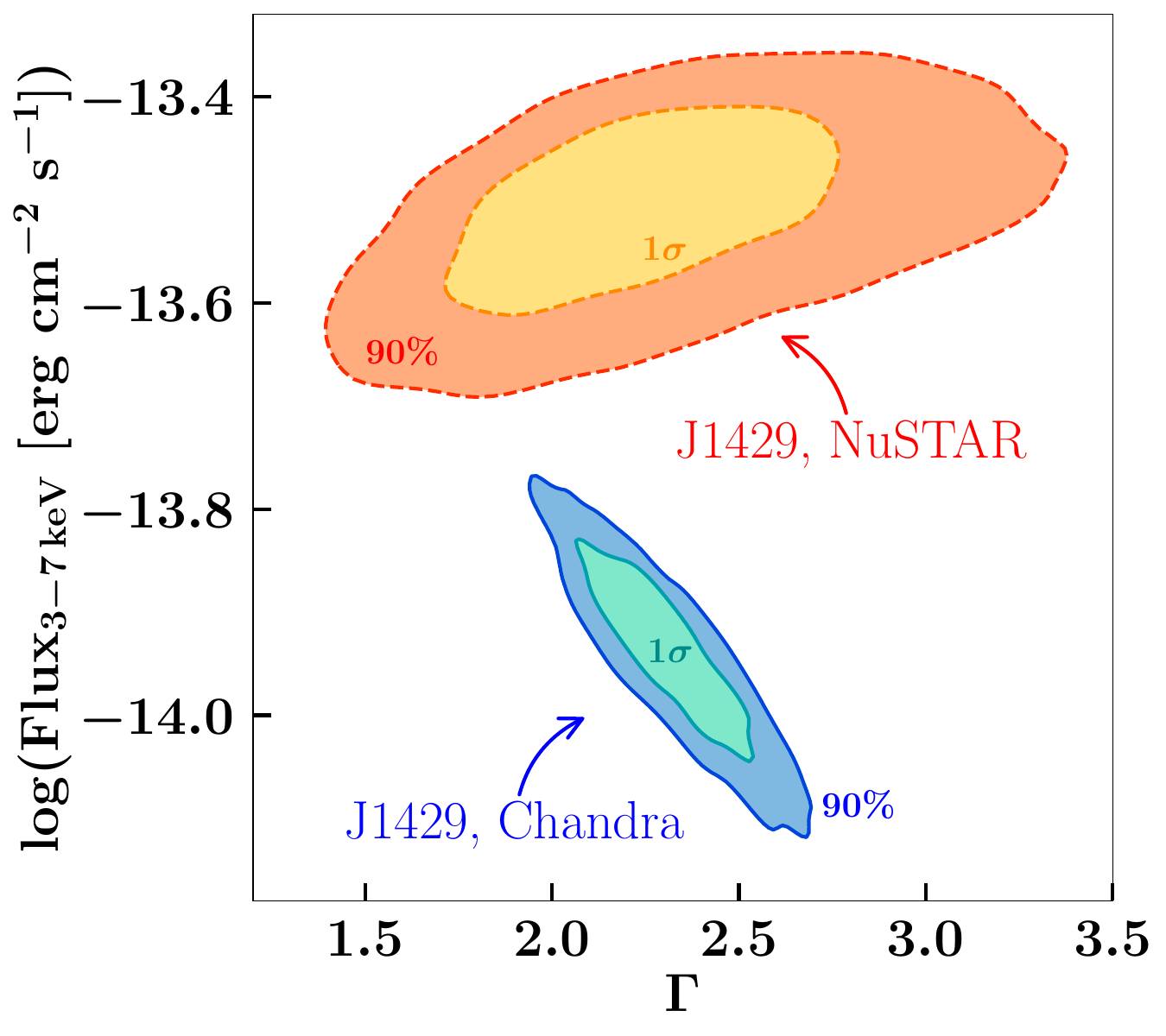}
    \caption{$3-7\,\rm keV$ posterior flux distribution versus the full-band photon index posterior distribution for \shortname\ obtained with the simultaneous fit of the three \textit{Chandra} sources and the co-added {\it NuSTAR} FPMAs (details in Section~\ref{sec:month-time}). The $1\sigma$ and 90\% contour levels are reported in the plot. There is no overlap between the two contours, which is a further indication that the flux of the source varied in the four months (observed frame) between observations. The photon index distributions span a similar range, indicating that more likely the source only varied in flux. Note that these photon indices are extracted by fitting datasets covering different energy ranges, and hence cannot be compared one-to-one.}
    \label{fig:dist_flux}
\end{figure*}

The results of our fit are presented in Table~\ref{tab:spectralfits}; we find that \shortname\ increased in flux by roughly 0.4 dex between observations. We show the background-subtracted unfolded spectra, best fits, and uncertainties in Figure \ref{fig:x-ray-fit}. To further demonstrate the extent of the flux change, we plot the posterior distributions of $\Gamma$ and ${\rm F}_{\rm 3\textrm{--}7~keV}$ in Figure \ref{fig:dist_flux}. For both Figures \ref{fig:x-ray-fit} and \ref{fig:dist_flux}, we stress that the \textit{NuSTAR} posteriors obtained for \shortname\ depend on the posteriors of all other \textit{Chandra} components. 


\begin{figure*}[t]
    \centering
    \includegraphics{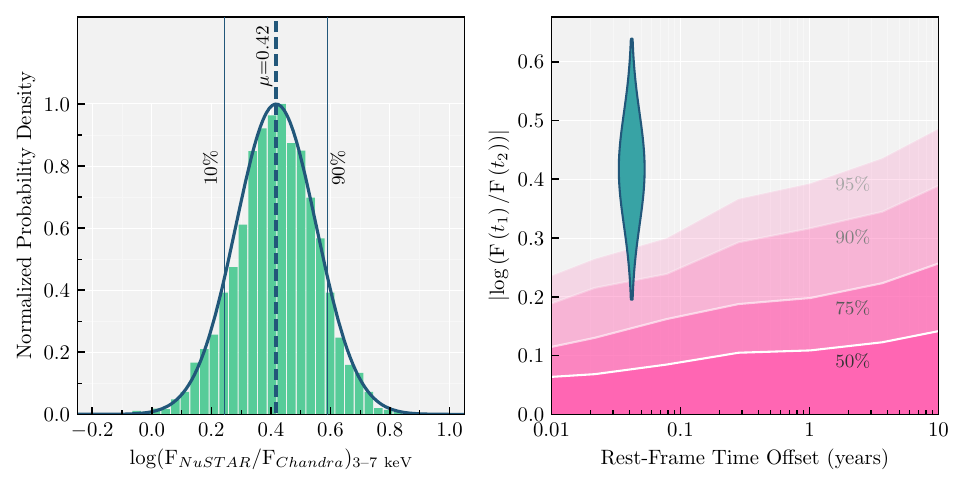}  
    \caption{{\bf Left:} Histogram of the difference between the \textit{Chandra} and \textit{NuSTAR} $3-7\,\rm keV$ posterior flux distribution obtained by simultaneously fitting all three \textit{Chandra} data sets and the \textit{NuSTAR} co-added FPMA. The mean (dashed vertical line) and 90\% confidence limit (CL, solid vertical lines) of the distributions obtained by a Gaussian fit (solid line) are shown. This histogram highlights that \shortname\ varied in flux between the \textit{Chandra} and \textit{NuSTAR} epoch of observation at $>90\%$ CL. The mean of the distribution indicates that the \textit{NuSTAR} flux is $\sim2.6$ times higher than the \textit{Chandra} one. Moreover, only 0.26\% of the distribution has a value $\leq0$, in support of the fact that it is highly unlikely that the source did not vary between the two epochs. \textbf{Right:} 90\% distribution of flux offsets observed for \shortname\ (dark blue-green) relative to flux offsets observed for lower-redshift ($z=[1,2]$) AGN in a broadly similar regime of black hole masses ($M_{\rm BH}>10^8$) and Eddington ratios ($\lambda_{\rm Edd} > 0.01$) by \citet[from dark to light pink]{2024MNRAS.531.4524G}. Most of the probability distribution of observed brightening occupies the region where fewer than 5\% of comparison flux variations reside. The rapidity of such a significant flux offset is thus unusual for an AGN.}
    \label{fig:flux_var}
\end{figure*}

The posterior distributions shown in Figure \ref{fig:dist_flux} do not overlap within their 90\% confidence intervals, indicating that the flux varied at a statistically significant level. But not only do the two distributions not overlap, the value of flux they encompass does not overlap, either\textemdash such that they do not allow for a scenario in which the flux remained constant while the spectral shape varied. To better demonstrate this effect, we plot the distribution of flux offsets (in effect marginalizing over the values of $\Gamma$) in Figure \ref{fig:flux_var}. We find an average logarithmic flux increase in the 3\textendash 7 keV band, $\log \left({\rm F}_{NuSTAR} / {\rm F}_{Chandra} \right)$, of 0.42, with a one-tailed 90\% lower limit of 0.24. The magnitude of this offset relative to typical AGN behavior is discussed in Section \ref{sec:disc-AGN}.

\subsection{More Complex Models}\label{sec:highecut_bknpo}

The harder energies probed by the \textit{NuSTAR} observations mean that, in comparison to previous \textit{XMM-Newton} or \textit{Chandra} observations, we could potentially observe more complex features arising in the observed spectra beyond a simple power law. For example, for X-ray emission produced solely by the AGN corona, we would expect to see a coronal cut-off at high energies\textemdash potentially observable by \textit{NuSTAR} \citep[e.g.][]{Ricci_2017, balokovic_2020, Kammoun_2024}. Such cut-offs are markers of the nature of the X-ray emitting particles and coronal physics. Conversely, the soft energies may be biased, such as by a Compton hump, with the harder energies better tracing the coronal emission \citep[e.g.][]{TA_2021,2024FrASS..1135459B}.  Alternatively, if the emission is coming from a jet, we could see a softening at higher energies in the spectrum if we were sampling the peak and/or the high-energy tail of the electron energy distribution producing the visible emission (radiating either via external Compton or synchrotron self-Compton processes; e.g., \citealp{An_2018, An_2020, Marcotulli_2020, Marcotulli_2022, Moretti_2021, 2022A&A...663A.147S, Gokus_2024}).

Owing to the fact that BXA enables us to sample the full parameter space for different models, and hence to discern which part of the parameter space is allowed given our data, we tested these more complex models by fitting \shortname\ with
\begin{enumerate}
    \item a redshifted, broken power law (M3; Section~\ref{sec:mod}) and
    \item a redshifted power law with an exponential cut-off (M4; Section~\ref{sec:mod}).
\end{enumerate}
As done for the simple power law case (Section~\ref{sec:month-time}), \textit{Chandra} spectra of I-SW and I-NE were also included in the fit.

Our results show that the fits of the observed spectra to these more complex models are not favored by the data (see Appendix~\ref{apx:highecut_bknpo} for details on the spectral fit results). In particular, the photon indices found through these fits are consistent with that of the simple power law case (M1). Nevertheless, the posterior distributions of these fits enable us to establish a lower-limit to the location of the possible spectral cut-off, $E_{\rm cut, obs-frame}\geq23.6\,\rm keV$, or equivalently $E'_{\rm cut, rest-frame} \geq 169\,\rm keV$.
Such a value is only slightly constraining given the luminosity of the source \citep[e.g.,][]{2019MNRAS.484.2735M}. However, the uncertainties in the fit parameters and residuals do not indicate an improvement with respect to the power-law fit.

\section{Discussion}
\label{sec:Discussion}

The detection of \shortname\ with eROSITA \citep{2020MNRAS.497.1842M} marked not just one of the most distant AGN detected in X-rays \citep[e.g.,][]{2016ApJ...823L..37A, 2018A&A...614A.121N, 2018ApJ...856L..25B, 2024arXiv240707236B, 2019ApJ...887..171C, 2020ApJ...900..189C, 2022ApJ...924L..25Y}, 
    but also the revelation that it was, by far, the most X-ray luminous source in the Epoch of Reionization. With a reported $L_{2-10\,\rm keV}=2.6^{+1.7}_{-1.0}\times10^{46}\,\rm erg\,s^{-1}$, it is almost an order of magnitude brighter than other runner up $z>6$ quasars \citep[e.g.][]{Belladitta_2020,2021ApJ...922L..24C}. Subsequent \textit{XMM-Newton} observations confirmed this luminosity and constrained the source's spectral parameters \citep{2021MNRAS.504..576M}, while additional high angular resolution \textit{Chandra} observations provided additional insights, including the detection of two nearby X-ray sources unresolved by the original observation \citep{2023MNRAS.524.1087M}. In this work we analyzed the first \textit{NuSTAR} observations of this system; the hard energy spectrum is well described by a power law of $\Gamma_{3\text{--}78\ {\rm keV}}=2.22^{+0.60}_{-0.54}$, consistent with the observed \textit{XMM-Newton} and \textit{Chandra} spectral properties of the source. Importantly, we find to more than 90\% confidence that the source flux varied over the observed four month baseline between \textit{Chandra} and \textit{NuSTAR} observations, at a level of ${\rm F}_{NuSTAR}/{\rm F}_{Chandra} = 2.6^{+1.0}_{-0.7}$. The derived \textit{NuSTAR} broadband luminosity of the source in the $21.54-71.8\,\rm keV$ rest frame of the source is $L_{21.54-71.8\,\rm keV}= 1.7_{-0.2}^{+0.5}\times10^{46}\,\rm erg~s^{-1}$; in the canonical $2-10\,\rm keV$ rest-frame, the quasar luminosity increased from $L_{2-10\,\rm keV,\rm Chan.}= (1.6\pm0.4)\times10^{46}\,\rm erg~s^{-1}$ to $L_{2-10\,\rm keV, NuS.}= 4.1^{+2.2}_{-1.6}\times10^{46}\,\rm erg~s^{-1}$.

These results raise three questions, which we discuss in turn: what is the scale of the variability for \shortname, what could have caused such a rapid, intense increase in the X-ray flux, and what future insights are needed to better understand this and similar sources?

\subsection{Scale of variability}

\begin{figure*}[t]
    \centering
    \includegraphics[width=0.8\textwidth]{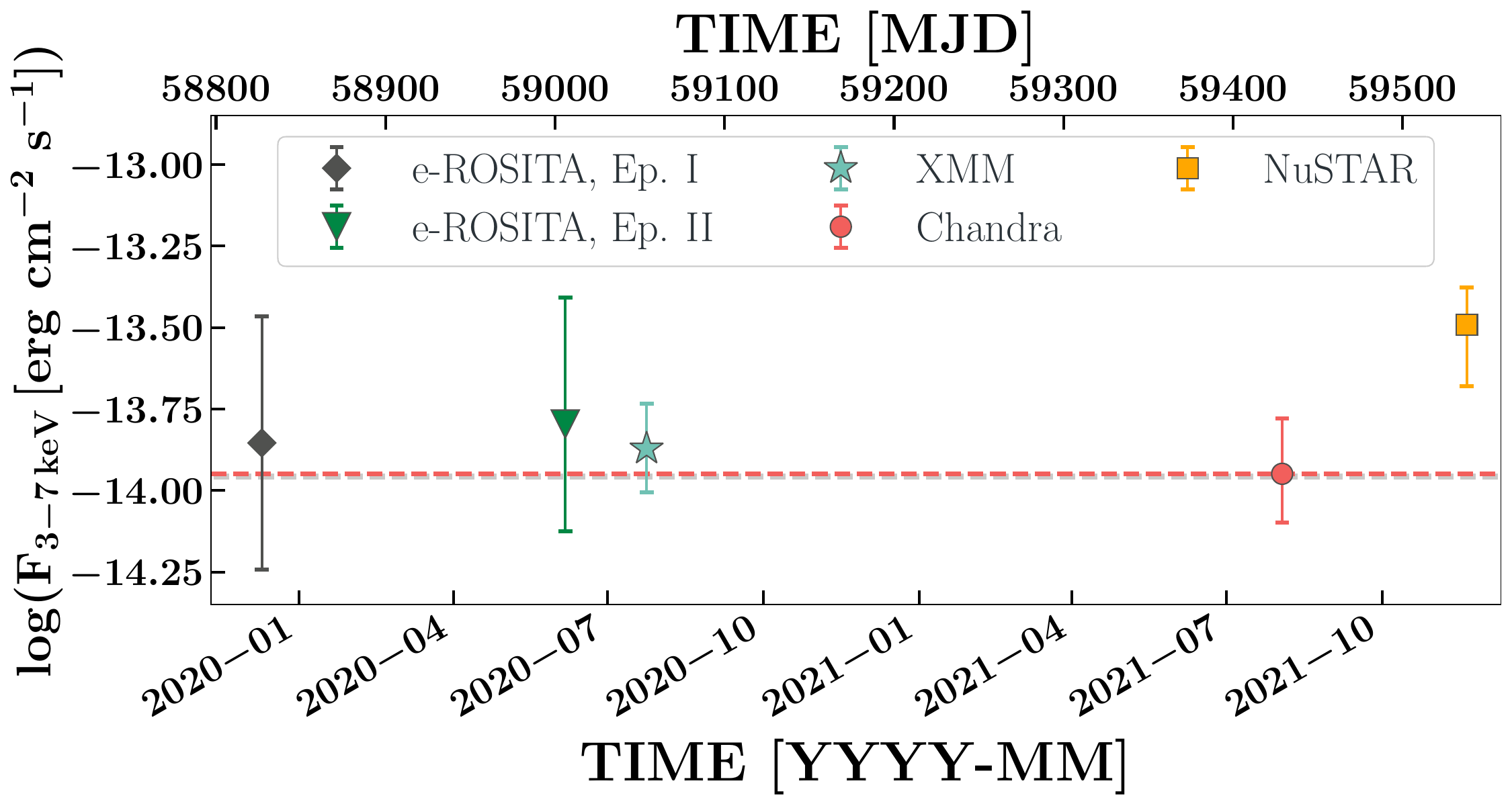}    
    \caption{$3-7\,\rm keV$ light-curve of \shortname; the errors are the 90\% HDI levels. The eROSITA flux points (black diamond \& green triangle) are derived by extrapolating the spectrum to the $3-7\,\rm keV$ range using $\Gamma=2.5$ and the eROSITA $0.2-6\,\rm keV$ fluxes reported in \citet{2021MNRAS.504..576M}. The \textit{XMM-Newton} flux point (cyan star) was derived using the best-fit power law model of \citet{2021MNRAS.504..576M}.  
    The \textit{Chandra} (red dot) and \textit{NuSTAR} (yellow square) data points are the ones obtained in this work. To guide the eye, the red dashed line at the level of the \textit{Chandra} flux is plotted. This plot highlights that there was no evidence of variability for \shortname\ before the \textit{NuSTAR} observation,
    when the source flux brightened by 0.4 dex. }
    \label{fig:x-ray_lightcurve}
\end{figure*}

The \textit{NuSTAR} observations presented here are now the fifth reported X-ray flux measurements for \shortname, following two eROSITA observations (\citealp{2020MNRAS.497.1842M} and \citealp{2021MNRAS.504..576M}), the \textit{XMM-Newton} follow-up \citep{2021MNRAS.504..576M}, and the \textit{Chandra} visit analyzed here \citep{2023MNRAS.524.1087M}. There are also several other repeated observations in other wavebands that enable a check on the type of variability \shortname\ displays.

To properly discuss these X-ray fluxes, we first convert previous measurements into the common 3\textendash 7 keV band. \citet{2021MNRAS.504..576M} provide flux values for the two eROSITA observations: ${\rm F}_{0.2\text{--}6\ {\rm keV}} = 1.1^{+0.6}_{-0.5}\times 10^{-14}$ and $1.3\pm 0.6 \times 10^{-14}\ {\rm erg}\ {\rm s}^{-1}\ {\rm cm}^{-2}$ for epochs \Romannum{1} and \Romannum{2}, respectively. These values were extrapolated using the best-fitting model of \citet{2021MNRAS.504..576M}, with $\Gamma=2.5$, which we, in turn, use to extrapolate flux values in our desired band. We find a first epoch flux of ${\rm F}_{3\text{--}7\ {\rm keV}}^{\rm \bf e\Romannum{1}} = (1.4\pm0.7) \times 10^{-14}\ {\rm erg}\ {\rm s}^{-1}\ {\rm cm}^{-2}$ and second epoch flux ${\rm F}_{3\text{--}7\ {\rm keV}}^{\rm \bf e\Romannum{2}} = (1.6\pm0.7) \times 10^{-14}\ {\rm erg}\ {\rm s}^{-1}\ {\rm cm}^{-2}$. Likewise, we use the fits presented in \citet{2021MNRAS.504..576M} to re-derive a new \textit{XMM-Newton} flux of  ${\rm F}_{3\text{--}7\ {\rm keV}} = 1.3_{-0.2}^{+0.3}\times10^{-14}\,\rm erg~cm^{-2}~s^{-1}$.

We show these fluxes, as well as the newly measured \textit{Chandra} and \textit{NuSTAR} fluxes, in Figure~\ref{fig:x-ray_lightcurve}. We note that only these two new data points account for the contamination of the two interlopers, and that the eROSITA and \textit{XMM-Newton} points may thus be biased towards a higher flux value; indeed, although \textit{XMM-Newton} observations resolved and excluded I-SW, \citet{2021MNRAS.504..576M} used a spectral extraction region that includes I-NE \citep[see discussion in][regarding the effect of I-NE of the \textit{XMM-Newton} spectra]{2023MNRAS.524.1087M}. Nevertheless, aside from the \textit{NuSTAR} observation, we find that the quasar flux has been otherwise stable, within uncertainties. The intra-\textit{XMM-Newton} variability was tested by \citet{2021MNRAS.504..576M} and further explored by \citep{2023MNRAS.524.1087M} in the context of the {\it Chandra} observations, with neither study finding any statistically significant variation. As such, 
there is no evidence that the X-ray flux of \shortname\ varied significantly until the \textit{NuSTAR} observation analyzed here.


As a radio-loud quasar, \shortname\ has now been observed in three epochs by the Very Large Array Sky Survey \citep[VLASS,][]{2020PASP..132c5001L}. The measured 3 GHz flux densities are $2.01\pm 0.13$ mJy (2017 December 01), $1.71\pm0.12$ mJy (2020 September 01), and $1.95\pm0.12$ mJy (2023 February 10). Again, no significant variability in the radio intensity is seen; if the \textit{NuSTAR} excess is related to the radio emission, it is thus more likely to be of a flaring nature than of a continuous variation.

\subsection{Origin of the Rapid Variability}

\subsubsection{AGN Variability}
\label{sec:disc-AGN}
We first consider if this variability could be due to standard processes in AGN. Assuming coherent emission from a non-jetted AGN,
the time scale of the variability constrains the size of the reprocessing component:
\begin{equation}
    R \leq c \times \Delta t_{\rm rest-frame}.
\end{equation}
{For \shortname, the $\sim15$ rest-frame days offset} corresponds to $R=72\, R_{\rm Schw}\approx 0.01\ {\rm pc}$, where the Schwarzschild radius for \shortname\ is $R_{\rm Schw}\sim5\times10^{14}\,\rm cm$. 


The X-rays in typical AGN are produced via inverse Compton scattering of accretion disk photons by a hot plasma of electrons known as the AGN corona \citep[e.g.][]{Haardt_1991}. This primary X-ray spectrum, usually described by a power-law with an exponential cut-off, can also be reprocessed (and even absorbed) by material in the AGN surroundings\textemdash the accretion disk itself ($R_{\rm disk}=1-500\,\rm R_{\rm Schw}$), the broad line region ($R_{\rm BLR}\lesssim0.1\,\rm pc$), and the torus ($R_{\rm Tor}\lesssim10\,\rm pc$). 
Assuming this is an adequate description for \shortname, the short variability timescale indicates that the most likely reprocessing component of the X-ray emission is the inner accretion disk 
and that the observed flux enhancement could be caused by variability in the corona itself or in the accretion flow \citep[e.g.][]{Ponti_2012, Puccetti_2014, Tortosa_2023, Serafinelli_2024}. For example, if the accretion was (close to) super-Eddington, then timescale variation in the disk can happen on short timescales \citep[see Discussion in][]{2022A&A...663A.159V}. 

Considering the estimated lower limit on the luminosity of \shortname\ below 2 keV from \citet[$L_{<2\,\rm keV}=1.2\times10^{47}\,\rm erg~s^{-1}$]{2020MNRAS.497.1842M}  and the 2\textendash100 keV luminosity of the source from our newly derived fits, we estimate the bolometric luminosity of the quasar following \citet{2020MNRAS.497.1842M} to be
\begin{equation}
    L_{\rm bol} = L_{<2\,{\rm keV}} + L_{2\text{\textendash}100\, {\rm keV} }.
\end{equation}
We find $L_{\rm bol, Chan.}\sim1.4\times10^{47}\,\rm erg~s^{-1}$ and $L_{\rm bol, Nu.}\sim1.8\times10^{47}\,\rm erg~s^{-1}$. The black hole mass from \citet{2019ApJ...873...35S} gives us an Eddington luminosity of $L_{\rm Edd}=2.3\times10^{47}\,\rm erg~s^{-1}$; therefore we can also relate the X-ray variability to an increase in Eddington ratio from $\log(\lambda_{\rm Edd})\sim -0.2$ to $\log(\lambda_{\rm Edd})\sim-0.1$, values consistent with the high-end tail of typical AGN Eddington rates \citep[e.g.][]{Suh_2015}. Hence, the accretion regime in the system\textemdash already quite fast\textemdash could have been slightly enhanced, leading to the observed increase of X-ray flux. 

To put \shortname\ in context with other AGN, in the right panel of Figure~\ref{fig:flux_var} we compare the observed flux variability amplitude and timescale of \shortname\ with the observed variability seen in a broad sample of AGN, as compiled by \citet{2024MNRAS.531.4524G}. To compare to a sample of similar sources, we limit this analysis to black hole masses of $M_{\rm BH} > 10^8\ {\rm M}_\odot$ and Eddington ratios of $\log(\lambda_{\rm Edd}) > -2$; we also require at least five times the source counts than the expected background for each measurement. Although that work samples the  0.2\textendash 2.3 keV energy range, the self-similar nature of power laws means that logarithmic flux offsets will be consistent with our chosen 3.0\textendash 10 keV band. As can be seen in Figure \ref{fig:flux_var}, the observed variability is extreme for that seen in similar AGN at similar time scales; if the flux variation comes from typical AGN processes, it would represent a significant statistical outlier.

 One of the key results presented by \citet{2024MNRAS.531.4524G} is that the amplitude of flux variation decreases with increasing mass and accretion rate; as such, not only is this comparison sample thus better suited than others that lack mass and Eddington ratio information (e.g., \citealt{2020MNRAS.498.4033T}, which shows stronger typical flux variations), but, by extending more than a decade below \shortname\ in mass and Eddington ratio, the sample is also conservative. While both the \citet{2024MNRAS.531.4524G} and \citet{2020MNRAS.498.4033T} analyses focus on radio-quiet quasars, \shortname\ is radio-loud; although the expected amplitude of flux variations may be larger for radio-loud quasars, these variations are often attributed to jets or beaming \citep{2017MNRAS.472.3789C}\textemdash topics we address below\textemdash rather than from the AGN and its corona.

The spectral index derived by our analysis is $\Gamma_{\it NuSTAR}=2.22_{-0.54}^{+0.60}$, fully consistent with typical X-ray coronae in the local universe ($\Gamma=1.8-2.0$, e.g.~\citealp{Ricci_2017}). 
Softer spectral behavior has been seen for AGN at the Epoch of Reionization \citep[e.g.,][]{Vito_2019, 2021ApJ...908...53W, Zappacosta_2023}, hinting to either an evolution of intrinsic coronal spectral properties with redshift or a lower coronal temperature regulated by a hybrid thermal/non-thermal plasma \citep[e.g.][]{Coppi_1999}.
 Ideally, if we were indeed looking at the emission from an unobscured AGN, 
we would like to constrain the cut-off of the corona, often used to infer the coronal temperature.
Our fit gives us a lower limit on the high-energy cut-off, $E_{\rm cut\text{-}off, rest\text{-}frame}>169\,\rm keV$, and an X-ray luminosity in the $2-10\,\rm keV$ band that increased from $L_{Chandra}= (1.6\pm0.4)\times10^{46}\,\rm erg~s^{-1}$ to $L_{NuSTAR}= 4.1^{+2.2}_{-1.6}\times10^{46}\,\rm erg~s^{-1}$. Our lower limit on the coronal cut-off energy agrees with the theoretical expectation \citep[see][]{2015MNRAS.451.4375F, 2019ApJ...875L..20L} for a $10^9 M_{\odot}$ black hole with luminosities of \shortname\ in either the slab or hemispheric coronal scenario, as well as the observational properties of other local AGN \citep[where coronal cut-offs have been measured in the range $E_{\rm cut-off,rest-frame}\gtrsim100\textendash 500\,\rm keV$; e.g.][]{2018MNRAS.480.1819R,balokovic_2020, Akylas_2021}. If present, it is likely that the turn-over happens in the spectral range dominated by the \textit{NuSTAR} background ($\geq 20\text{\textendash}30\,\rm keV$, observer frame), although this is currently unconstrained by our data quality. 


An extreme possibility to consider is for the detected variability to have an obscuration-based origin. Indeed, Compton-thick BLR and/or outflow-based obscuration has been proposed to be more common at higher redshifts \citep[e.g.][]{Bertola_2020,Gupta_2022,maiolino_2024}. However, for the rest-frame $\gtrsim$\,20\,keV continuum of J1429 to have increased by a factor of $>2$ would require an enormous change in column density on the order of 
$\Delta N_{\rm H}\gtrsim$\,10$^{24}$\,cm$^{-2}$ (see Figure~1 of \citealt{Boorman_2024_nulands}). Such extreme changes of Compton-thick column densities have been observed in the local universe (e.g.,
\citealt{Risaliti_2005,Marinucci_2016,Marchesi_2022}), as well as rapid changes in obscuration on the order of weeks (e.g., \citealt{Risaliti_2002,Elvis_2004,TA_2023}); nevertheless, occurrences of both are drastically rarer (e.g., \citealt{Ricci_2023}). Lastly, for typical Compton-thick column densities, the true intrinsic unobscured rest-frame 2\,--\,10\,keV luminosity would be $\sim$\,1.5\,--\,3 orders of magnitude higher, which would be remarkably high.

\subsubsection{Blazar}
\label{sec:disc-blazar}
An alternative explanation for many of the properties of \shortname, including the rapid variability presented here, is that \shortname\ is a blazar\textemdash that is, that we are seeing it along the jet axis. This interpretation would resolve tensions in many of the source's properties, chief of which would be the extreme luminosity, as blazars are widely observed to be X-ray luminous, especially at high redshifts \citep{2019MNRAS.489.2732I}. Additionally, \citet{2011A&A...531L...5F} note that \shortname\ hosts a jet that is only 100 pc long in projection, despite the AGN being luminous and radio-loud; if seen face-on, however, the small size of this jet would merely be a projection effect \citep{2024ApJ...964...98X}. Blazar variability is observed to be bursty \citep[e.g.,][]{2024A&A...682A.100E}, in keeping with our observed dramatic brightening over a short time period despite otherwise consistent flux (Figure \ref{fig:x-ray_lightcurve}), and the compressed observational timescale of events seen along the jet \citep[e.g.,][]{Singh_2020} means that the emission region could be larger than assumed for the un-beamed case.

However, other observed properties of J1419+5447 are not consistent with blazars, such that \citet{2020MNRAS.497.1842M} previously rejected that hypothesis. As discussed above, the VLASS radio flux densities are relatively consistent between epochs, as are other radio intensity measurements \citep[see Table 1 in][]{2020MNRAS.497.1842M}. Similarly, the spectral slope of the radio emission is $\alpha\approx {-0.7}\ {\rm to}\ {-1.0}$ 
\citep{2011A&A...531L...5F,2017MNRAS.467.2039C}, classified as a steep-spectrum source, while blazars have typical values of $\alpha> -0.5$ \citep[e.g.,][]{2019MNRAS.490.5798D}. \citet{2019MNRAS.489.2732I} proposed a classification diagram for potential high-redshift blazars, requiring $\Gamma \lesssim 1.8$ and UV-to-X-ray slope\footnote{$\tilde{\alpha}_{\rm OX} = -0.3026\log(L_{\rm 10\,keV}/L_{\rm 2500\AA})$} $\tilde{\alpha}_{\rm OX} < 1.355$; although \shortname\ matches the latter ($\tilde{\alpha}_{\rm OX} = 0.96^{+0.11}_{-0.07}$, \citealp{2020MNRAS.497.1842M}), the steep X-ray slope seen in all analyses so far is inconsistent with this expectation. 

Of course, the properties of blazars are not always so easily confined, particularly at high redshift; indeed, multiple works now have reported that sources classified as blazars at high-redshift ($z>4$) display contradictory qualities \citep[i.e. classified as blazars with X-rays, but displaying misaligned/bent jets in radio;][]{Sbarrato_2015,Sbarrato_2021,2022A&A...663A.147S, Cao_2017, Caccianiga_2019, 2019MNRAS.489.2732I}. Moreover, some $z>5$ blazars detected in X-rays display softer photon indices ($\Gamma>2$; \citealp{An_2018,An_2020}) which has been interpreted as the peak of the synchrotron self-Compton component emerging at these energies. 
Furthermore, under the assumption \shortname\ may be a progenitor for one of the powerful MeV blazars observed at lower redshifts, the average MeV blazar SED shape derived from \citet{Marcotulli_2022} implies a rest-frame peak of a $ L_{\rm X}\sim10^{46}\,\rm erg~s^{-1}$ source at $E_{b}\sim 4\,\rm MeV$ and an average photon index of $\Gamma_{\rm X}=1.70\pm0.14$, which is compatible to our derived \textit{NuSTAR} and \textit{Chandra} photon indices within a 90\% significance level (see Figure~\ref{fig:dist_flux}).
Even more compelling for our case, \citet{2022AA...664A..39K} found that the radio spectrum of \shortname\ is flat ($\alpha\sim-0.5$) below 1.4 GHz and shows similar properties (including lack of radio variability) to the $z = 6.10$ blazar source J0309+2717 \citep{Belladitta_2020, Ighina_2022}.

 Given NOEMA observations showing evidence for significant jet impact on the quasar host's gas \citep{2024ApJ...962..119L}, it could be possible that the jet is being bent outside of the innermost accretion regions. As one potential manifestation of bending on our understanding of this source, if the face-on nature of the jet only emerges well away from the central engine, the much weaker magnetic fields may not produce the expected radio signatures of a typical blazar.

In the case that the X-ray jet is pointed at us, beaming would shorten the observed variability timescale \citep[e.g.,][]{Singh_2020} by 
\begin{equation}
    t_{\rm rest}=t_{\rm obs}\times\delta/(1+z),
\end{equation}
where the Doppler factor ($\delta$) is 
\begin{equation}
    \delta=\left(\Gamma_{\rm b}(1-\beta\cos\theta_{\rm V})\right)^{-1}
\end{equation}
for viewing angle $\theta_{\rm V}$, emission region speed $\beta$, and bulk Lorentz factor $\Gamma_{\rm b}$. For a typical value of $\Gamma_{\rm b}{\sim}5-10$ \citep[e.g.][]{Marcotulli_2020, Sbarrato_2021} in the limit of $\theta_{\rm V}\rightarrow0$, $t_{\rm source}=t_{\rm obs}\times\Gamma/(1+z)\sim 2.7-5.5\,\rm months$. Using the same formalism as above, this would imply that the size of the X-ray emission region is larger: $R=390-796\, R_{\rm Schw}$ ($0.07-0.1\,\rm pc$), placing it within the scale of the BLR region.

For a blazar jet, there are multiple ways that short-term multiwavelength variability could be produced, including a shock \citep[e.g.][]{Baring_2017}, magnetic reconnection \citep[e.g.][]{Zhang_2022}, or an increase or modification of the photon field surrounding the jet \citep[see][and references therein for recent reviews on the topic]{Blandford_2019, Bottcher_2019, Hovatta_2019}. The photon field could also be modified by a perturbation in the accretion disk \citep[e.g.][]{Stern_2018} that propagates in the BLR clouds, or perhaps by a passage of a cloud in front of the jet, thereby acting as a mirror to enhance the radiation field \citep[e.g.][]{Ghisellini_1996, LeonTavares2013, Chavushyan_2020}. 

A blazar interpretation for \shortname\ would be exciting, as the detection of one blazar implies that a larger population of radio-loud AGN with similar properties\textemdash but not favorably aligned to our line of sight\textemdash exist at a similar redshift \citep[e.g.,][]{Belladitta_2020}. Indeed, recent work has started to show evidence that a significant fraction of AGN in the early Universe are jetted \citep[e.g.,][]{2022A&A...663A.147S,2024arXiv240707236B}, such that the large number of known quasars should, in turn, imply a larger population of jetted quasars than currently seen. 
From the recent luminosity function work of \citet{Marcotulli_2022}, in the redshift bin $z=[6.0,6.3]$ and within the highest luminosity range $L_{\rm 14-195\,\rm keV}=[10^{47}-10^{48}]\,\rm erg~s^{-1}$, we would expect $N\lesssim 10$ blazars in an all-sky sample. Although this number can be thought of as an upper limit, being an extrapolation of the X-ray luminosity function, it is not improbable that radio-loud, X-ray luminous \shortname\ is indeed one of the expected blazars at those redshifts.

\subsubsection{Jets}
\label{sec:disc-jet}
Between the AGN and blazar interpretations is one in which the X-ray luminosity is being powered by jets. In particular, \citet{2021MNRAS.504..576M} advanced this hypothesis, as did \citet{2023MNRAS.524.1087M}, based on the possible presence of a second or extended component. The fundamental concept of this emission is that the energy density of the CMB, which scales as $\left( 1 + z \right)^4$, dominates over magnetic fields in these jets, and so most of their emission is in inverse Compton upscattered X-rays instead of radio \citep[e.g.,][]{2014MNRAS.438.2694G}. Direct evidence for this IC/CMB emission around $z\gtrsim6$ quasars has recently been seen \citep{2021ApJ...911..120C,Ighina_2022}, and for \shortname\, the presence of large jets visible only in X-rays would rectify the shortness of the observed radio structures with the independent NOEMA arguments for active, powerful jets. 

However, the rapid variability seen here contradicts this interpretation. The size of the X-ray producing regions of jets are significantly larger than the sub-pc-scale region permitted by the observed variation. \citet{2023MNRAS.524.1087M} already constrained the location of the bulk of the X-ray emission to be at $<3\,\rm kpc$ and emphasized that, to explain the level of \shortname's X-ray luminosity, Doppler boosting would still be required in an IC/CMB jet scenario. For the significant originally observed luminosity of \shortname\ to be due to non-beamed X-ray IC/CMB emission and for the observed short-term excess to be produced by a different mechanism, then the scale of the variation would have to be even more extreme. As such, the results of this work imply that the brightness of \shortname\ is not due to extended, unbeamed emission from its jets. 

\subsubsection{Interloper Contamination}
\label{sec:disc-int}
We must also consider the least exciting option, namely that the observed increase in flux came from one of the two interloper sources. In this case, \shortname\ did not vary in flux, but, instead, either I-NE or I-SW (or both) increased by a factor of 10 in brightness in around 50 days in their rest frames; the \textit{NuSTAR} image, with its worse angular resolution, masked the origin of the increase in flux. This scenario is excluded by design in our modeling, due to the following reasons.

\begin{enumerate}
    \item One decade in flux in such a short time scale would be a massive outlier in typical AGN variability (compare to Figure \ref{fig:flux_var} and see Appendix~\ref{sec:interlopers}), even more so than the already impressive value assumed in this work. Yet, unlike \shortname, the two interloper sources are not otherwise typical of sources that would produce such extreme flux variations; neither have associated radio sources, nor do they have X-ray spectra typical of blazars. Moreover, unlike \shortname, these sources are not already outliers in X-ray emission. As such, we would not expect them to produce such a massive outburst.
    \item Although the \textit{NuSTAR} observation suffers from relatively poor angular resolution, and so the source is smeared over the position of the quasar and interlopers, the centroid of the source nevertheless aligns with that of \shortname\ (Figure \ref{fig:sky_images}). \textit{NuSTAR} has an astrometric accuracy of roughly $8^{\prime\prime}$ to 90\% \citep{2013ApJ...770..103H}, and so offsets of $30^{\prime\prime}$ (I-NE) or $45^{\prime\prime}$ (I-SW) would be significant outliers. Conversely, there is an unrelated X-ray source, \objectname{J142922.1+544428}, seen at the edge of the \textit{NuSTAR} detectors located at roughly the expected astrometric precision.
\end{enumerate}

\subsection{Future Outlook}
For \shortname\textemdash as well as for many other quasars in the $z\gtrsim6$ epoch\textemdash both multiwavelength and coordinated observations are necessary to understand the nature of the source \citep[e.g.,][]{2024NatAs.tmp..293B, 2024ApJ...977L..46B}. Based on the variability seen here, the radio properties, which are key to determining the possibility of the source being a blazar, need to be coordinated to guard against flux variation inducing an incorrect measurement of the spectral shape, while simultaneity with X-ray observations is critical to assess jet conditions. Likewise, IR imaging and spectroscopy, and sub-mm imaging can constrain the stellar and SMBH parameters, providing the needed context of the host galaxy, its merger environment, and how physically impactful the jets are.


Given the significant exposure time required to detect \shortname\ with \textit{NuSTAR}, and given that \shortname\ is, even with variability, the most X-ray luminous $z\gtrsim6$ quasar, it is likely that this will be the most distant object ever robustly detected by the observatory. Nevertheless, it is an indication of what will be possible with a future, more sophisticated hard X-ray telescope such as the proposed probe-class \textit{High Energy X-ray Probe} \citep[\textit{HEX-P};][]{2024FrASS..1157834M}. Although \textit{HEX-P} was not selected in the 2023 Astrophysics Probe Explorer opportunity, the challenges that motivate it still remain. The broadband X-ray coverage (0.2\textendash 80 keV) proposed for \textit{HEX-P} would enable simultaneous soft-to-hard measurements, constraining the broadband high-energy spectrum to levels beyond current capabilities (at both soft and hard energies)\textemdash in significantly shorter exposure times. Moreover, the high sensitivity and low background at high energies would allow us to not only detect countless more $z\gtrsim6$ quasars beyond 5 keV but to fit more complex and informative models to this population, such as first measurements of their coronal breaks. The full capabilities of \textit{HEX-P} on blazar and AGN science were explored in detail by \citet{Marcotulli_2024}, \citet{2024FrASS..1135459B}, \citet{Kammoun_2024} and \citet{Pfeifle_2024}, but this pathfinder observation with \textit{NuSTAR} demonstrates that high-redshift science can be an important part of the next \textit{HEX-P} caliber mission's capabilities.

\section{Summary}

In this work, we have presented \textit{NuSTAR} observations of \shortname, a $z=6.19$ radio-loud quasar that is the most X-ray luminous quasar currently known in the early universe. These observations establish a new record for the most distant object observed by \textit{NuSTAR} (surpassing blazars B2 1023+25 at $z=5.3$ and QSO J0906+6930 at $z = 5.48$; \citealp{2013ApJ...777..147S,An_2018}) and demonstrate the capabilities of observing hard-energy X-rays into the Epoch of Reionization.

Owing to an unrelated \textit{Chandra} observation four months prior, we were able to fit the quasar's X-ray spectrum while self-consistently accounting for contamination from two nearby sources. We found that, over just two weeks in the quasar rest-frame between the \textit{Chandra} and \textit{NuSTAR} observations, the flux of the source increased by 0.4 dex. 

This observed X-ray variability is one of the most extreme seen at these redshifts, and it is potentially indicative of significant events in either the quasar accretion region or the jet. Further observations, some of which are already underway, are needed to further establish the nature of these fluctuations.

\vspace{5mm}
{\small 
We thank the journal anonymous referee for the 
constructive review.
Support for this work was provided by the National Aeronautics and Space Administration through Chandra Award Number GO3-24069X, GO8-19093X, and GO0-21101X issued by the \textit{Chandra} X-ray Center, which is operated by the Smithsonian Astrophysical Observatory for and on behalf of the National Aeronautics Space Administration under contract NAS8-03060.

This research has made use of data from the NuSTAR mission, a project led by the California Institute of Technology, managed by the Jet Propulsion Laboratory, and funded by the National Aeronautics and Space Administration. Data analysis was performed using the \textit{NuSTAR} Data Analysis Software (NuSTARDAS), jointly developed by the ASI Science Data Center (SSDC, Italy) and the California Institute of Technology (USA). The scientific results reported in this article are based on observations made by the \textit{Chandra X-ray Observatory} contained in \dataset[CDC 308]{https://doi.org/10.25574/cdc.308}. 
This research has made use of software provided by the \textit{Chandra} X-ray Center (CXC) in the application package CIAO. 

LM acknowledges that support for this work was provided by NASA through the NASA Hubble Fellowship grant No. HST-HF2-51486.001-A awarded by the Space Telescope Science Institute, which is operated by the Association of Universities for Research in Astronomy, Inc., for NASA, under contract NAS5-26555.
T.C. and A.S. acknowledge support from NASA Contract NAS8-03060 to the Chandra X-ray Center.
Portions of T.C.'s research were supported by an appointment to the NASA Postdoctoral Program at the Jet Propulsion Laboratory, California Institute of Technology, administered by Universities Space Research Association under contract with NASA. The National Radio Astronomy Observatory is a facility of the National Science Foundation operated under cooperative agreement by Associated Universities, Inc. 
G.M. acknowledges financial support from the INAF mini-grant "The high-energy view of jets and transients" (Bando Ricerca Fondamentale INAF 2022).
CM acknowledges support from ANID BASAL project FB210003. 
YK thanks the support of the German Space Agency (DLR) through the program LEGACY 50OR2303. Research at the Naval Research Laboratory is supported by NASA DPR S-15633-Y.

}


\vspace{5mm}
\facilities{NuSTAR, CXO}

\software{
BXA \citep{Buchner_2014},
CIAO \citep{2006SPIE.6270E..1VF},
PyFITS \citep{1999ASPC..172..483B}, 
XSPEC \citep{1996ASPC..101...17A}
          }

\appendix

\section{Cross-Calibration Test}\label{sec:cross_cal_appb}
An important factor to take into account in our analysis is the effect of cross-calibration constants between different instruments. Specifically, \citet[][cf.~Table 6 in their work]{Madsen_2017} showed that for a simultaneous \textit{NuSTAR}/\textit{Chandra} observation of a non-variable soft power-law spectrum ($\Gamma\sim2.5$) source, if the constants of FPMA/B are frozen to 1, then the constants assigned to the \textit{Chandra} ACIS spectra should be 1.09/1.10. In other words, due to calibration systematics, the \textit{Chandra}-derived flux is systematically $\sim10\%$ higher than the \textit{NuSTAR} one. 

To test whether this factor would make any difference to our analysis, we repeated the fit detailed in Section~\ref{sec:month-time} assigning (and freezing) the value of 1.09 to the cross-calibration constant of the three \textit{Chandra} spectra (\shortname, I-NE, I-SW), while keeping the cross-calibration constant of the co-added FPMAs frozen to unity. The results show that making this assumption of systematics in the cross-calibration returns an even stronger variability of \shortname. 
This finding, combined with the fact that the most updated \textit{NuSTAR} calibration files have been re-calibrated so that the derived \textit{NuSTAR} flux is 5-15\% higher\footnote{See Calibration page entry 2021-10-26: \url{https://nustarsoc.caltech.edu/NuSTAR_Public/NuSTAROperationSite/software_calibration.php}} \citep[hence should be comparable to the \textit{Chandra} flux, see][]{Madsen_2022}, validates not just keeping the cross-normalization frozen to unity for all instruments, 
but also that this choice ultimately does not impact our findings \citep[see also discussion in Sec.~5.3 of][]{Sobolewska_2023}.

\section{High-Energy Cutoff and Broken Power Law scenarios}\label{apx:highecut_bknpo}
\begin{figure*}[h!]
    \centering
    \includegraphics[width=0.85\textwidth]{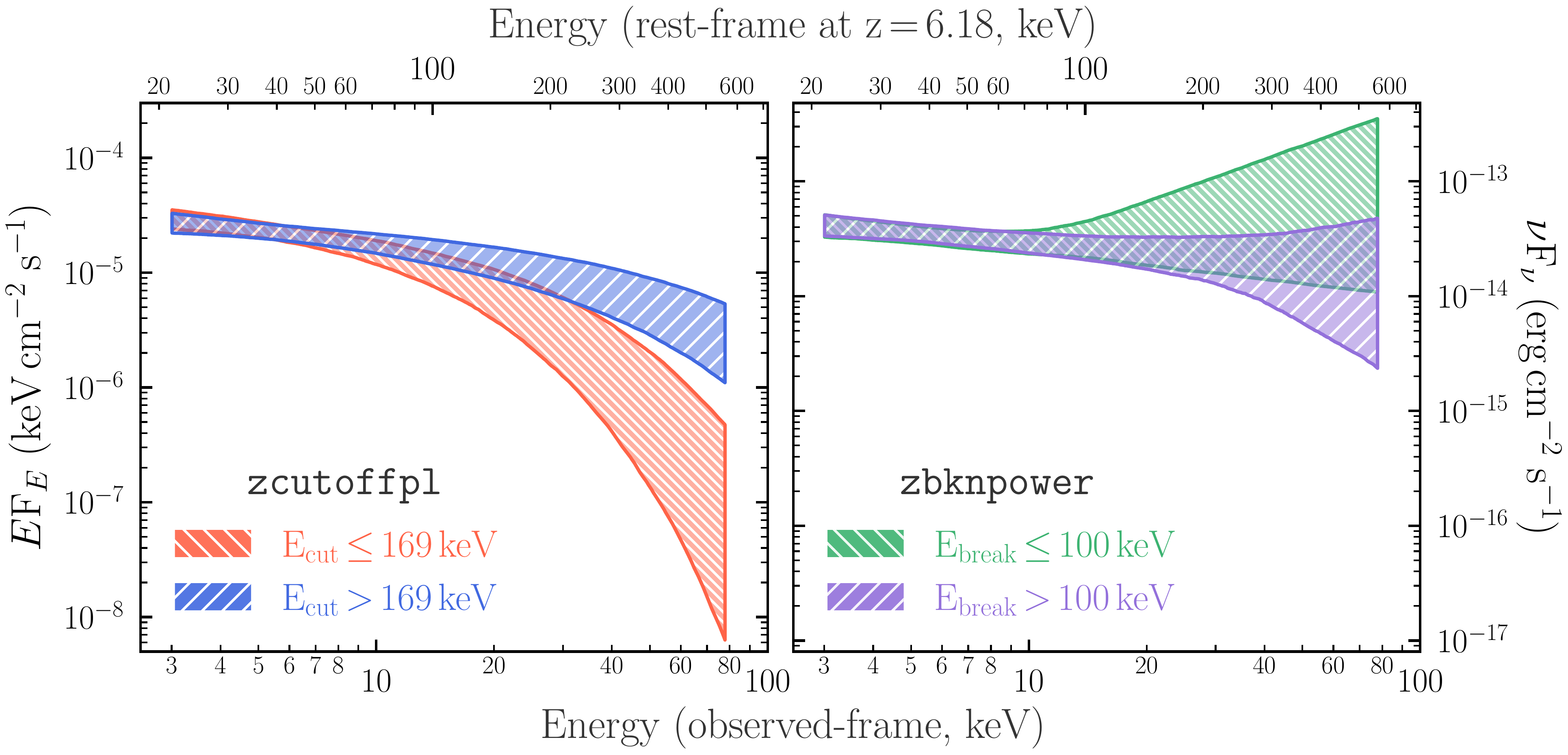}
    \caption{$3-78\,\rm keV$ spectrum of \shortname\ using a more complex spectral model. {\bf Left}: redshifted power law with high-energy cutoff. The shaded bands in the plot represent the 68th percentile range regions of model predictions as calculated considering all posteriors for which $E_{\rm cut, rest-frame}<169\,\rm keV$ (orange, backward-slashed shaded region) and $E_{\rm cut, rest-frame}>169\,\rm keV$ (blue, forward-slashed shaded region, i.e.~akin to the simple power-law model). {\bf Right}: redshifted broken power law. The shaded bands in the plot represent the 68th percentile range regions of model predictions as calculated considering all posteriors for which $E_{\rm b, rest-frame}\leq100\,\rm keV$ (green, backward-slashed shaded region, i.e. region of the posterior that includes values of $E_{\rm b}$ at $68\%$ HDI level from the mode of the distribution) and $E_{\rm b, rest-frame}>100\,\rm keV$ (purple, forward-slashed shaded region, i.e.~akin to the simple power-law model). These plots show that, within uncertainties, we cannot determine whether a complex modeling of the source is favored. Therefore, the best-fit representation of the data at hand is the simplest one.}
    \label{fig:bkn_highecut_case}
\end{figure*}

In our analysis, we conclude that the observed spectrum of \shortname\ is well explained by a power law. However, we also tested more complex\textemdash yet physically motivated\textemdash shapes to test if \shortname\ shows a break or curvature in the spectrum (see Section~\ref{sec:highecut_bknpo} for details on the fitting procedure). We use both a power-law with exponential cut-off (M4) and a broken power-law (M3) to interpret the data. 
The power law with exponential cutoff gives us an upper limit 
on the cut-off energy value of $E_{\rm cut\text{-}off, rest\text{-}frame}>169\,\rm keV$ and $\Gamma=2.15_{-0.19}^{+0.24}$. The broken power law gives us 
$\Gamma_1=2.34_{-0.27}^{+0.26}$, $\Gamma_2=2.21_{-2.14}^{+0.37}$ and $\log(E_{\rm break, rest-frame}, \rm keV)=1.88_{-0.16}^{+0.93}$.
Within uncertainties, all the scenarios are consistent with the simple power-law case. In Figure~\ref{fig:bkn_highecut_case} we showcase the spectral posterior distribution of these two models in the 3\textendash 78 keV band. It can be seen that in either case, a broken/cutoff spectrum is fully consistent (within the 68\% percentile range) with the power-law scenario. 

\section{Interlopers Posterior Flux and Photon Index distribution}\label{sec:interlopers}

\begin{figure*}[h]
    \centering
    \includegraphics[width=0.75\textwidth]{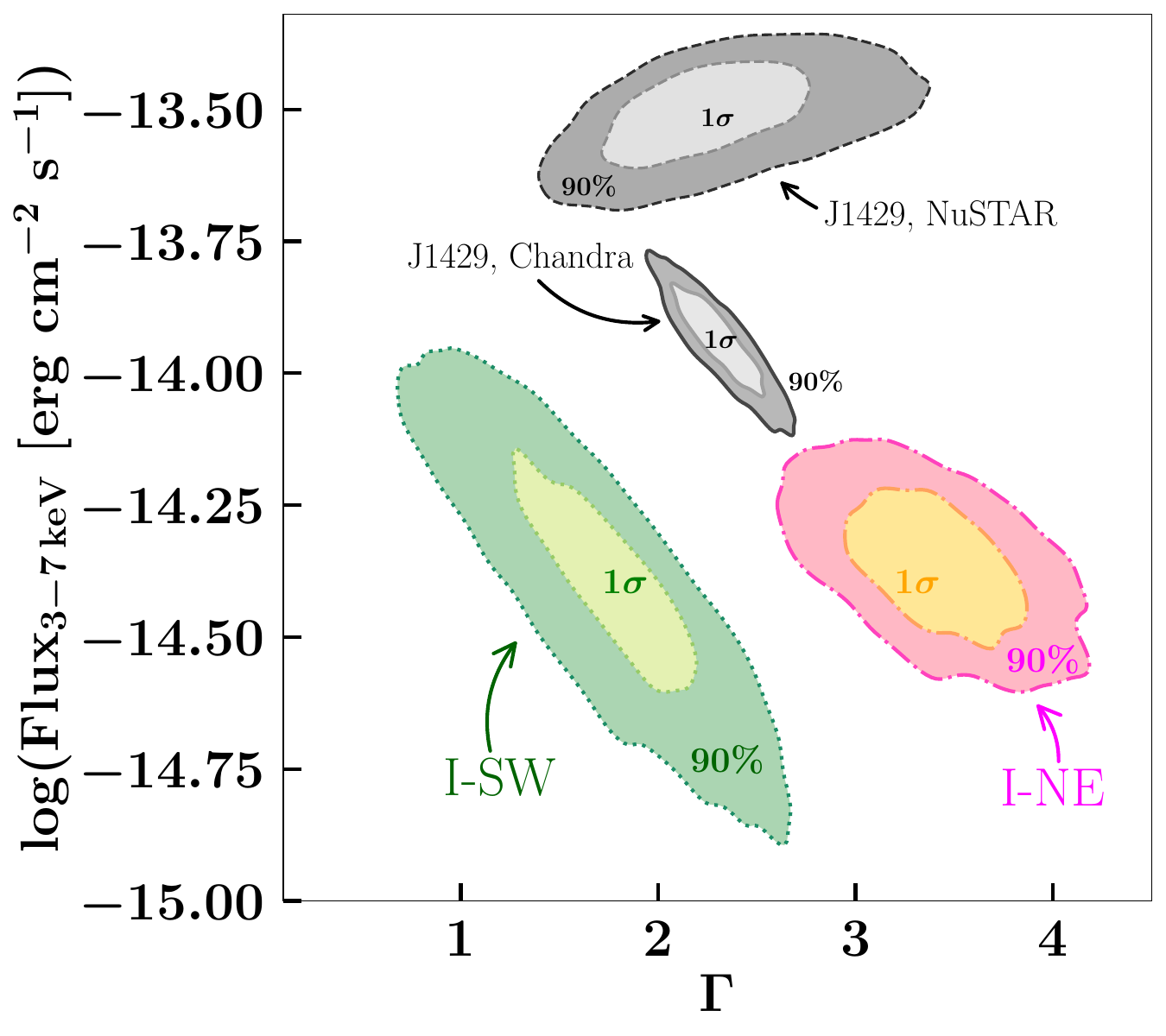}
    \caption{$3-7\,\rm keV$ posterior flux distribution versus the full-band photon index posterior distribution for \shortname\ (\textit{Chandra} epoch: solid lines; \textit{NuSTAR} epoch: dashed lines), I-NE (dotted lines) and I-SW (dash-dotted lines) obtained with the simultaneous fit of the 3 \textit{Chandra} sources and the co-added FPMAs (details in Section~\ref{sec:month-time}). The $1\sigma$ and 90\% contour levels are reported in the plot. This is an extension of Figure~\ref{fig:dist_flux} to show how subdominant I-NE and I-SW are in the $3-7\,\rm keV$ band compared to \shortname.}
    \label{fig:all_post}
\end{figure*}

The combined fit detailed in Section~\ref{sec:month-time} returns the posterior probability on all fitted parameters and flux distributions for all the loaded spectra. For completeness, we show in Figure~\ref{fig:all_post} the version of Figure~\ref{fig:dist_flux} including the interlopers. This Figure highlights how subdominant the interlopers are with respect to \shortname. As noted in Section~\ref{sec:disc-int}, the interlopers would have needed to vary approximately by a decade in brightness in around 50 days rest-frame to recover the \textit{NuSTAR} flux reported in this work. 

\bibliography{references}{}
\bibliographystyle{yahapj}

\end{document}